\documentclass[prodmode,acmtoms]{acmsmall}

\usepackage{amsmath}
\usepackage{amssymb}
\usepackage{stmaryrd}
\usepackage{amscd}
\usepackage{tikz}
\usetikzlibrary{matrix,arrows,decorations.pathmorphing,decorations.pathreplacing,positioning}
\usepackage{mathtools}
\usepackage{listings}
\usepackage{upquote}
\usepackage{listings}
\usepackage{subfig}
\usepackage{booktabs}
\usepackage{hyperref}
\usepackage[ruled]{algorithm2e}
\renewcommand{\algorithmcfname}{ALGORITHM}
\SetAlFnt{\small}
\SetAlCapFnt{\small}
\SetAlCapNameFnt{\small}
\IncMargin{-\parindent}

\definecolor{LightBlue}{rgb}{0.8,0.8,0.8}
\definecolor{MidBlue}{rgb}{0.4,0.4,0.4}
\definecolor{DarkBlue}{rgb}{0.00,0.00,0.55}
\definecolor{Black}{rgb}{0.00,0.00,0.00}
\hypersetup{
    linkcolor = DarkBlue,
    anchorcolor = DarkBlue,
    citecolor = DarkBlue,
    filecolor = DarkBlue,
    urlcolor = DarkBlue,
    linkcolor = DarkBlue,
    anchorcolor = Black,
    citecolor = DarkBlue,
    filecolor = Black,
    urlcolor = DarkBlue,
    colorlinks  = true,
    pdfauthor   = {Simon Funke <s.funke09@imperial.ac.uk>},
    pdftitle    = {A high-level framework for PDE-constrained optimisation}
}

\definecolor{dkgreen}{rgb}{0,0.6,0}
\definecolor{gray}{rgb}{0.5,0.5,0.5}
\definecolor{mauve}{rgb}{0.58,0,0.82}
\definecolor{lightyellow}{rgb}{255,254,230}

\newcommand{\fenics}{{\mbox{FEniCS}}\xspace}
\newcommand{\dolfin}{{\mbox{DOLFIN}}\xspace}

\newcommand{\da}{\mbox{{dolfin-adjoint}}\xspace}

\newcommand{\libadjoint}{\mbox{{libadjoint}}\xspace}

\newcommand{\ponedgptwo}{\mbox{P1$_\textrm{DG}$-P2}\xspace}
\newcommand{\ponecg}{\mbox{P1$_\textrm{CG}$}\xspace}
\newcommand{\pzerodg}{\mbox{P0$_\textrm{DG}$}\xspace}
\newcommand{\ptwopone}{\mbox{P2-P1}\xspace}

\newcommand{\dx}{\,\mathrm{d}x}
\newcommand{\dt}{\,\mathrm{d}t}

\begin{document}
\markboth{S. W. Funke and P. E. Farrell}{A framework for automated PDE-constrained optimisation}

\title{A framework for automated PDE-constrained optimisation}
\author{S. W. Funke and P. E. Farrell\affil{Imperial College London}}

\begin{abstract}
A generic framework for the solution of PDE-constrained optimisation problems based on the \fenics system is presented.
Its main features are an intuitive mathematical interface, a high degree of automation, and an efficient implementation of the generated adjoint model.
The framework is based upon the extension of a domain-specific language for variational problems to cleanly express complex optimisation problems in a compact, high-level syntax. 
For example, optimisation problems constrained by the time-dependent Navier-Stokes
equations can be written in tens of lines of code. 
Based on this high-level representation, the framework derives the associated adjoint equations in the same domain-specific language, and uses the FEniCS code generation technology to emit parallel optimised low-level C++ code for the solution of the forward and adjoint systems. 
The functional and gradient information so computed is then passed to the optimisation algorithm to update the parameter values.
This approach works both for steady-state as well as transient, and for linear as well as nonlinear governing PDEs and a wide range of functionals and control parameters.
We demonstrate the applicability and efficiency of this approach on classical textbook optimisation problems and advanced examples.

\end{abstract}




\acmformat{Funke, S. W. and Farrell, P. E. 2013. A framework for automated PDE-constrained optimisation.} 

\begin{bottomstuff}
This work is supported by the Grantham Institute for Climate Change, a Fujitsu CASE studentship, EPSRC grant EP/I00405X/1, and a Center of Excellence grant from the Research
Council of Norway to the Center for Biomedical Computing at Simula Research Laboratory. 
The authors would like to acknowledge helpful discussions with D. A. Ham and M. E. Rognes, 
the contribution of T. Surowiec to the example presented in section \ref{sec:mpec}, and 
A. S. Candy and A. Avdis for the help with the mesh generation in section \ref{sec:turbine_optimisation}.

Author's address: Funke and Farrell: Department of Earth Science and Engineering, Imperial College London. 
Funke: Grantham Institute for Climate Change, Imperial College London.
Farrell: Center for Biomedical Computing, Simula Research Laboratory, Oslo.
\end{bottomstuff}

\maketitle

\section{Introduction}\label{sec:introduction}

Optimisation problems constrained by partial differential equations (PDEs) are ubiquitous across science
and engineering. Such problems consist of optimising an objective functional, e.g. maximising the performance or
minimising the cost of a system, subject to constraints given by the laws of physics \cite{lions1971}: for example,
an aeronautical engineer will want to choose the best shape for a wing to optimise its performance
\cite{jameson1988}. Inverse problems may also be treated as optimisation problems, where
the goal is to infer some unobservable state from observable evidence; this is achieved by adjusting
the unknown state to minimise some misfit functional \cite{ledimet1986}. This approach is now
fundamental to the geosciences: for example, it is routinely used in operational meteorology \cite{rabier2000}.

Approximating the solution of PDEs is computationally expensive. This motivates the use of gradient-based optimisation
algorithms, since exploring the control space without derivative information typically requires a
prohibitive number of PDE evaluations for practical problems.
The typical case for PDE-constrained optimisation problems is that where the dimension of the
control space is large, and where the number of functionals to control is small (usually one).
Therefore, their efficient solution relies on the fast gradient computation for a small number of functionals with respect to many parameters.

A na\"ive approach to computing the gradient of a functional is to perturb each control
parameter in turn, and approximate the gradient using finite differences. A more sophisticated way would be to employ the
tangent linear model associated with the forward PDE system, which circumvents the problems of
roundoff errors by propagating derivative information forward through the computational graph, from one
input parameter through to all outputs. However, with both of these approaches, the number of PDE
solves required for a single gradient computation scales linearly with the number of parameters,
making them infeasible for the typical case described above. By contrast, the adjoint method
computes the gradient of a scalar functional with a single PDE solve, by propagating derivative information backwards
through the computational graph, from the output functional back to all inputs
\cite{giles2000,griewank2008}.  The adjoint method is a key ingredient in making the large-scale
solution of complex optimisation problems feasible.

However, deriving and implementing the adjoint PDE model is typically regarded as
difficult. This has been one of the main motivations for the development of algorithmic
differentiation techniques (AD, also called automatic differentiation), which attempt to automate
the adjoint model derivation. However, in practice the application of an AD tool typically requires large amounts of user intervention and expertise,
in particular for advanced forward model implementations.  \citeN[pg.
\emph{xii}]{naumann2011} states that ``the automatic generation of optimal (in terms of robustness
and efficiency) adjoint versions of large-scale simulation code is one of the great open challenges
in the field of High-Performance Scientific Computing''. \citeN{giles2000} observe that 
\begin{quote}
Considering the importance of design to .. all of engineering, it is perhaps surprising that the
development of adjoint codes has not been more rapid .. [I]t seems likely that part of the reason is
its complexity. 
\end{quote}

In previous work, we have efficiently solved this adjoint derivation problem for the case where the
forward problem may be discretised using the finite element method \cite{farrell2012}. The key contribution of
this paper is to apply this advance to automate the solution of large classes of
PDE-constrained optimisation problems.
The main features of this new framework are:
\begin{description}
 \item[Usability] The user specifies the discretised optimisation problem in a form that resembles the mathematical notation. 
 \item[Automation] Based on this problem specification, the framework performs the necessary steps for the optimisation, without further user intervention.
These steps include interfacing with an optimisation method, followed by repeated PDE solves for
evaluating the objective functional and computing the functional gradient using the automatically derived adjoint system.
\item[Generality] The framework handles large classes of governing PDEs, including coupled, nonlinear and time-dependent PDEs.
Furthermore, the user may choose from multiple gradient-free and gradient-based optimisation algorithms.
 \item[Performance] Optimisation algorithms typically require many iterations, each of which involve computationally expensive PDE solves.
For many problems of practical interest, efficient parallel PDE solvers are therefore essential to
obtain reasonable run times. 
Significant work has been undertaken in order to produce efficient assembly code in \fenics~\cite{kirby2007b,markall2012,olgaard2010}.
Since the adjoint model relies on the same \fenics code generation techniques as the forward model, 
the performance of the adjoint implementation inherits the same efficiency and parallel scaling of the forward model.
As a consequence, the achieved adjoint efficiency ratios are often close to the optimal ratio between 1 and 2; by
contrast, a general AD tool typically yields efficiency ratios in the range of 3 to 30 \cite[pg. \emph{xi}]{naumann2011}.
\end{description}
These features are achieved by exploiting the particular structure of finite element models (in
contrast to traditional AD, which attempts to solve the general case). Finite element discretisations of the
governing PDE have a natural domain-specific language, the language of variational forms, that
abstractly captures the mathematical structure of the problem without any details of how its
solution is to be achieved on a particular platform.  Instead of implementing the model in tens or
hundreds of thousands of lines of a low-level language such as Fortran or C++, the user compactly
describes the discretisation of the forward problem in the Unified Form Language
(UFL) \cite{alnaes2011,alnaes2012} of the FEniCS project
\cite{logg2011}. UFL mimics the mathematical notation almost exactly, and can express even complex
time-dependent coupled nonlinear problems in tens of lines of code. 

The presented framework uses the high-level UFL representation of the forward model for two purposes. 
First, the forward code is generated using the \fenics project in the usual way: the UFL is passed to a dedicated finite element compiler, which
\emph{generates} optimised low-level C++ code for its parallel implementation on a particular platform
\cite{kirby2006,olgaard2010,markall2012}. Second, as each forward solve is executed at runtime,
a symbolic tape of the forward model is recorded in UFL format. This tape is analogous to the concept of
a tape in AD \cite{corliss1993}, except that instead of recording individual floating-point operations, the units on the tape are
whole equation solves. Once the forward model has terminated, symbolic manipulation is
applied to this tape to derive the UFL representation of the associated adjoint problem, which in turn is
passed to the same compiler to emit an efficient parallel implementation of the adjoint model~\cite{farrell2012}. 

In this paper we discuss the extension of this system to compactly express and solve optimisation problems. 
The user describes the forward model, the control parameters and the objective functional in an extension of the UFL notation. 
The optimisation framework then repeatedly 
re-executes the tape to evaluate the functional value, solves the adjoint PDE to compute the functional gradient, 
and modifies the tape to update the position in parameter space until an optimal solution is found.

\subsection{Related work}\label{sec:related_frameworks}
One of the main motivations for this work is the fact that, despite the broad applicability of PDE-constrained optimisation, 
there exist few software packages that gather and unify the tools required to solve such problems.

A closely related project is developed by \citeN{waanders2002}, with the goal of creating an optimisation framework based on the finite-element software Sundance~\cite{long2012}.
Sundance is similar to \fenics in that it also operates on variational forms: in particular, it can automatically differentiate and adjoin individual variational forms.  
However, the built-in automatic adjoint derivation of Sundance does not currently extend to cases where the forward model consists of a sequence of variational problems, 
which is typically the case for time-dependent problems.

Other open source alternatives include DAKOTA \cite{eldred1996}, the Stanford University Unstructured suite \cite{palacios2012} and RoDoBo \cite{becker2005}.
The key difference between these and the framework presented here is that the difficult step of the adjoint derivation and implementation has not been automated. 
Instead, if a new PDE is to be solved the adjoint model must be derived and implemented, either manually or with the help of AD tools,
both of which demand significant development effort and user expertise.

Finally, the PROPT~\cite{propt}, ACADO~\cite{houska2011} and CasADi~\cite{andersson2012} toolkits are optimisation frameworks with similar design goals, but focussed on ordinary differential equations and differential-algebraic equations instead of PDEs.

The paper is organized as follows: 
The next section states the general form of PDE-constrained optimisation problems and compares different solution techniques.
Section~\ref{sec:the_optimisation_framework} presents the newly developed framework in detail, followed by a code demonstration in section~\ref{sec:examples}.
In section~\ref{sec:verification} the framework implementation is verified using textbook examples with known analytical solutions. 
Finally, section~\ref{sec:applications} applies the optimisation framework to a wide range of problems before making some concluding remarks in section~\ref{sec:summary}. 

\section{The formulation of PDE-constrained optimisation problems}\label{sec:optpde_problem_formulation}

We consider the PDE-constrained optimisation problem in the following general form:
\begin{equation}
\begin{aligned}
 \min_{ u,  m}~&J( u,  m) \\ 
 \textrm{subject to }  
& F(u, m)  = 0, \\
& h( m)  = 0, \\ 
& g( m)  \le 0, 
\end{aligned}\label{eq:general_optimisation_problem}
\end{equation} % Hinze page 57
where the vector $m$ contains the optimisation parameters,
$F( u,  m) = 0$ is a system of PDEs parameterised by $m$ with solution vector $u$, and
$J( u,  m) \in \mathbb{R}$ is the scalar-valued objective functional that is to be minimised.
The equality and inequality constraints $h(m) = 0 $ and $g(m) \le 0 $ enforce additional conditions  on the optimisation parameters $m$.
A common example is a box constraint of the form:
\begin{equation*}
   a \le  m \le  b.
\end{equation*}
State constraints are not directly considered in formulation~\eqref{eq:general_optimisation_problem}. 
However, in section~\ref{sec:mpec} we show an example where a penalisation approach is employed to enforce state constraints.

%Inequality constraints on the state solution $u$ are not considered in this work: they a very recent development in the context of PDE-constraint optimisation, and 
%are typically enforced using regularisation techniques and semismooth Newton methods~\cite[\S2.7]{hinze2009}.
Throughout this paper we assume that the above operators are sufficiently differentiable, and that the PDE has a unique solution for any $m$, i.e.
there is a solution operator $u(m)$ such that $F(u(m), m) = 0 \quad \forall \ m$.

\subsection{The optimality conditions and the reduced formulation}\label{sec:the_optimality_system}
The necessary first order optimality conditions for problem~\eqref{eq:general_optimisation_problem}, also known as the Karush-Kuhn-Tucker (KKT) conditions~\cite{karush1939,kuhn1951},
are derived from the associated Lagrangian.
Ignoring the control constraints for simplicity, the Lagrangian of~\eqref{eq:general_optimisation_problem} is:
\begin{equation*}
 \mathcal L(u, m, \lambda) \equiv J(u, m) + \lambda^T F(u, m) \in \mathbb R,
\end{equation*}
where $\lambda$ is the Lagrange multiplier (also known as the dual or adjoint variable).
The KKT conditions state that for every local minimum $(\bar u, \bar m)$ at which some regularity conditions are satisfied (see~\citeN{hinze2009} for details), there exists a Lagrange multiplier $\bar \lambda$ such that:
\begin{subequations}
 \begin{align}
  \frac{\partial \mathcal L}{\partial u}(\bar u, \bar m)& = \frac{\partial  J}{\partial u}(\bar u, \bar m)  + \bar \lambda^T \frac{\partial  F}{\partial u}(\bar u, \bar m) = 0, \label{eq:optimality_condition} \\  
  \frac{\partial \mathcal L}{\partial m}(\bar u, \bar m) & = \frac{\partial  J}{\partial m}(\bar u, \bar m)  + \bar \lambda^T \frac{\partial  F}{\partial m}(\bar u, \bar m) = 0, \label{eq:adjoint_equation} \\  
  \frac{\partial \mathcal L}{\partial \lambda}(\bar u, \bar m) & = F(\bar u, \bar m) = 0. \label{eq:state_equation} 
 \end{align}\label{eq:kkt_system}%
\end{subequations}
Equation~\eqref{eq:adjoint_equation} is referred to as the adjoint equation; \eqref{eq:state_equation} recovers the governing PDE.
Solving \eqref{eq:adjoint_equation} for $\bar \lambda$ and substituting the result into the control equation~\eqref{eq:optimality_condition} yields that the total derivative of the objective functional vanishes at the optimal point.

% The generalized reduced gradient / reduced eleminitation method is described in "Optimization Theory and Methods", Wenyu Sun, Ya-Xiang Yuan, section 11.2-11.3 (page 502pp):
% can be downloaded here: http://link.springer.com/book/10.1007/b106451/page/1
One approach to compute a local solution of problem~\eqref{eq:general_optimisation_problem},
known as the all-at-once approach, simultaneous analysis and design, or the oneshot approach, is to directly solve the KKT system~\eqref{eq:kkt_system}. 
A common solution method is sequential quadratic programming (SQP), which for the simplified case considered here is equivalent to applying Newton's method to the KKT system~\eqref{eq:kkt_system}.
A key advantage of SQP is that it inherits the fast local quadratic convergence to a local solution from Newton's method~\cite{boggs1996}.
However, the KKT system~\eqref{eq:kkt_system} yields significant challenges for numerical solvers:
it is a coupled, nonlinear and often ill-conditioned system of PDEs and the resulting linear systems that need to be solved for each SQP update are high-dimensional.
This issue becomes particularly problematic in the case of time-dependent governing PDEs where the discrete solution vectors $u$ and $\lambda$ contain the entire forward and adjoint solution trajectories over time and space.

Since direct solver methods are typically not suitable to solve linear problems of such dimensions, 
iterative solvers in combination with advanced preconditioning techniques are often applied and show promising results in certain applications~\cite{battermann2001,biros2000,schoeberl2007}.
An alternative approach is to use a space reduction method; here, the solution of the full linear systems is avoided 
by performing a block-LU decomposition~\cite{byrd1991,biegler1995,schulz1998}. 
As a result, the system is uncoupled and can be solved in separate steps of more manageable size.
Reduced SQP methods have been successfully applied to various applications~\cite{kupfer1992,orozco1992,orozco1997}. 
%However, since these methods are relatively new, only few software packages are available, many of which are still in development.
%Parallelisation is crucial for good performance but non-trivial to implement in a generic way.

The all-at-once approach has the disadvantage that the current estimate of $u$, $m$ and $\lambda$ must be stored at any time. 
For large, time-dependent problems, storing the whole estimate of the forward and adjoint solutions $u$ and $\lambda$ can quickly exceed the available memory:
% Formula: 2*8*10^8*10^6/((1024^4) (forward + adjoint)
for example, a simulation with $10^6$ spatial degrees of freedom and $10^8$ timesteps would require roughly $1000$ TB of memory in double point precision, 
exceeding the memory capacities of most available computers.

This issue can be circumvented by performing a space reduction on the original optimisation problem~\eqref{eq:general_optimisation_problem}.
This approach (also known as black-box or nested analysis and design approach) replaces the objective functional $J(u, m)$ with the reduced functional~$\hat J(m) \equiv J(u(m), m)$, 
that is the functional is considered as a pure function of the optimisation parameters.
Since the reduced functional implicitly enforces the solution of the PDE, the PDE-constraint becomes superficial in the optimisation formulation.
The result is the following {reduced optimisation problem}:
\begin{equation}
\begin{aligned}
 \min_{ m} \hat J( m)& \\
\textrm{subject to } 
 h( m) & = 0, \\% \label{eq:reduced_general_optimisation_problem_equality_constraint} \\
     g(m) & \le 0. % \label{eq:reduced_general_optimisation_problem_unequality_constraint}
\end{aligned}\label{eq:reduced_general_optimisation_problem}%
\end{equation}

This formulation has the practical advantage that the dimension of the optimisation problem is greatly reduced, 
since the dimension of $m$ is typically much smaller than that of the PDE solution $u$. 
Consequently, many robust and established optimisation methods are directly applicable.
Furthermore, the storage requirement is significantly lowered;
firstly because the adjoint solution is only used for computing the functional gradient and does not need to be saved and
secondly because the storage of the entire forward solution trajectory may be avoided by using a checkpointing strategy to balance storage and computation cost. 
This allows the solution of large scale optimisation problems for which storing the whole forward solution would be impossible~\cite{griewank1992}.
% 1003 over 1000 is 167668501 which is larger than 10^8. The forward equations are evaluated three times instead of once. Hence (assuming that the forward and adjoint cost the same) we have as computational cost with checkpoint / without checkpoint the ratio:
% 3 + 1 / 1 + 1 = 2
In the example above, the optimal checkpointing scheme implemented by~\citeN{griewank2000} with $1000$ checkpoints reduces the storage cost to approximately $10$ GB 
while the computational cost approximately doubles.
Another advantage of the reduced formulation is that the governing PDE is always satisfied after each optimisation iteration.
Hence, the optimisation loop may be stopped as soon as the functional is sufficiently reduced, simplifying the formulation of termination criteria.
%This is a nonlinear optimisation algorithm with constraints, that can be solved with standard algorithms that are described in the next section.

\citeN{long2012} state that for steady problems, the all-at-once can outperform the reduced formulation by a wide margin, but that for time-dependent systems, the reduced formulation
is often preferable. Since the framework supports time-dependent problems, the current implementation solves the optimisation problem in the reduced formulation~\eqref{eq:reduced_general_optimisation_problem}.  However, in principle all components for solving the all-at-once approach are available, and we plan to implement this approach in future work. 
Furthermore, the reduced formulation is often used to precondition the all-at-once approach~\cite{biros2000}.

The following example illustrates the two formulations~\eqref{eq:general_optimisation_problem} and \eqref{eq:reduced_general_optimisation_problem} on a classical optimal control problem (see for example \citeN[chapter 2.1.5]{troeltzsch2005} or \citeN[chapter 1.5.3]{hinze2009}): 
given a thin, heatable plate~$\Omega \subset \mathbb R^2$ with fixed temperature at the domain boundary~$\partial \Omega$, what is the optimal heating function that should be applied to obtain a desired temperature 
profile?
This problem can be formulated as an optimisation problem constrained by the stationary heat equation:

\begin{equation}
\begin{aligned}
\min_{u, m}~&\frac{1}{2} \| u - u_d \|^2_{L^2(\Omega)} + \frac{\alpha}{2} \| m \|^2_{L^2({\Omega})} \\
\textrm{subject to} 
& -\nabla^2 u  = m \hspace{1cm} \textrm{on } \Omega, \\%\label{eq:optimal_control_with_heat_equation_state_equation_con_example} \\
& u  = 0 \hspace{2cm} \textrm{on } \partial \Omega, \\%\label{eq:optimal_control_with_heat_equation_state_equation_con_example_bc} \\
& a \le m \le b \hspace{1.3cm} \textrm{on } \Omega. 
\end{aligned}\label{eq:optimal_control_with_heat_equation_con_example}%
\end{equation}
Here $u_d: \Omega \to \mathbb R$ is the desired temperature profile and $u: \Omega \to \mathbb R$ is the solution of the stationary heat equation with homogeneous Dirichlet boundary conditions.
The objective functional measures the misfit between the PDE solution and the desired temperature profile plus a regularisation term that is multiplied by a scaling factor $\alpha  \ge 0$. 
The optimisation parameter $m: \Omega \to \mathbb R$ controls the heat source and is limited by the box constraints.

Under mild assumptions, the heat equation with Dirichlet boundary conditions yields a unique solution for any source parameter $m$~\cite[\S 1.3.1.1]{hinze2009}.
Hence there exists a solution operator $u(m)$ and the reduced problem may be formulated:
\begin{equation*}
\begin{aligned}
\min_m~&\frac{1}{2} \| u(m) - u_d \|^2_{L^2(\Omega)}  + \frac{\alpha}{2} \| m \|^2_{L^2({\Omega})}  \\
 \textrm{subject to } 
 & a  \le m \le b \hspace{1.5cm} \textrm{on } \Omega. 
\end{aligned}%\label{eq:optimal_control_with_heat_equation_con_example_reduced}%
\end{equation*}
This problem will be used for the code example in section~\ref{sec:examples}.

%The state system consists of a PDE that typically enforces the physical laws of the system. 
%The numerical solution of this PDE commonly involves the solution of large sparse matrix systems. 

\section{The optimisation framework}\label{sec:the_optimisation_framework}

The core of the framework relies on two software components: 
first, the \fenics system~\cite{logg2011,logg2007a} is used to solve the forward and adjoint PDEs.
Second, it relies on \libadjoint and \da~\cite{farrell2012} to automatically derive the associated adjoint system for gradient computations.
In this work we have extended the \da framework to go beyond adjoint derivation, to automate the solution of PDE-constrained optimisation problems.

The following sections discuss these components in detail.
The source code, including the examples and applications in the following sections, is available at \url{http://dolfin-adjoint.org}.

\subsection{The \fenics system} \label{sec:fenics}
The \fenics system is a collection of software components for automating
the solution of PDEs by the finite element method.
This section gives a brief introduction to the \fenics system.
A thorough overview can be found in~\citeN{logg2011}.

To solve a PDE with the \fenics system, the user defines its discretised weak form in the domain specific language UFL that mimics and encodes the mathematical formulation~\cite{alnaes2011,alnaes2012}.
This high-level formulation is then passed to a finite element form compiler such as FFC~\cite{kirby2006}, which generates optimised low-level code for the 
evaluation of the local element tensors.
This generated code is used by \dolfin~\cite{logg2010a,logg2011a} to globally assemble and solve the problem.
\dolfin also provides the data structures for meshes, function spaces and functions.

For time-dependent PDEs, the temporal discretisation is usually performed with a non-finite element discretisation scheme. 
In this case, the user writes the time loop manually and solves the variational problem for each timelevel as described above. 

\dolfin has interfaces for both C++ and Python.
The Python interface uses just in time compilation, i.e. it invokes the necessary compilers at runtime. 
In contrast, the code generation for the C++ interface happens at a preprocessing step before running the forward model. 
As a consequence, the high-level description of the forward problem is not directly available at runtime in the C++ interface. 
Because \da relies on this data to perform runtime inspection and manipulation, the remaining sections discuss only the Python interface to \dolfin. 

\subsection{\libadjoint and \da}
The libraries \libadjoint and \da enable the automatic derivation and solution of tangent linear and adjoint models from forward models written in \dolfin.

The purpose of \libadjoint is to facilitate the development of tangent linear and adjoint models based on the fundamental abstraction of considering the forward model as a sequence of equation solves. 
Based on this abstraction, the library builds a symbolic description of the forward model, the tape, from which it can automatically derive the symbolic representation of the associated tangent linear and adjoint systems. 

\begin{figure}
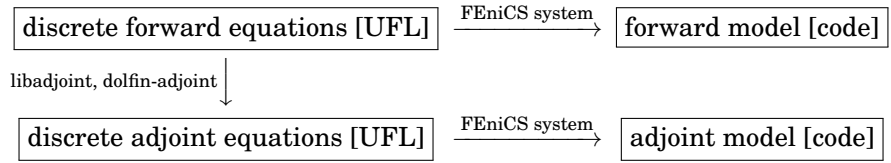

\begin{equation*}
  \begin{CD}
    \fbox{\textrm{discrete forward equations [UFL]}}
    @>\textrm{\fenics system}>>
    \fbox{\textrm{forward model [code]}}\\
    @V\textrm{\libadjoint, \da}VV
    \\
    \fbox{\textrm{discrete adjoint equations [UFL]}}
    @>\textrm{\fenics system}>>
    \fbox{\textrm{adjoint model [code]}}\\
  \end{CD}
\end{equation*}
  \caption{The automated generation of the adjoint model with \da.
    The user specifies the discrete forward equations in the high-level UFL language.
    From that symbolic representation of the problem, \libadjoint and \da can derive the corresponding representation of the discrete adjoint
    equations in UFL. 
    Both the forward and adjoint equations are passed to the \fenics system to generate the forward and adjoint model implementations.}
  \label{fig:our_approach} 
\end{figure}

The software library \da acts as the between \dolfin and \libadjoint. 
It inspects \dolfin's problem description at runtime, and performs the required tasks for applying \libadjoint without user intervention. 
The tangent linear and adjoint models produced with \libadjoint are represented in the same high-level data format as the forward model. 
Therefore, the code generation techniques in \fenics can be applied to the adjoint model in the same manner as the forward model, see figure~\ref{fig:our_approach}. 
\citeN{farrell2012} showed that this approach leads to a robust, automatic and efficient way of implementing tangent linear and adjoint models.  

\subsection{The optimisation framework}\label{sec:opt_user_interface}
\subsubsection{Introduction}

The proposed framework extends \da to solve PDE-constrained optimisation problems. 
The optimisation process consists of iteratively evaluating the functional of interest and its gradient at different points in the parameter space. 
The key idea is to automate these evaluations by operating purely on the tape of the forward model recorded by \libadjoint. 
In particular, a functional evaluation is obtained by replaying the tape (which runs the forward model) while simultaneously evaluating the objective functional. 
The functional gradient is computed by deriving and solving the adjoint model, as described in the previous section. 
When the optimisation algorithm updates the point in parameter space, the tape is modified accordingly. 

With this approach, the only inputs required from the user are the objective functional, the control parameters and the governing PDEs,
optionally with additional equality and inequality constraints.

\subsubsection{User interface}
The solver for an optimisation problem is typically implemented in the following three steps.

First, the user implements the governing PDE in the Python interface of \dolfin~\cite{logg2011}.
\dolfin supports linear and nonlinear as well as steady and transient PDEs, and has been used to solve complex coupled systems such
as the Reynolds-averaged Navier-Stokes equations for turbulent flows \cite{mortensen2011}, the Stokes equations for mantle convection \cite{vynnytska2013}, and
the Cahn-Hilliard equations for phase separation \cite{wells2006}.

Second, the user defines the objective functional and the optimisation parameters.
For that, a \texttt{Functional} class has been introduced that extends \dolfin to support the expression of time-dependent functionals.
For example, an objective functional computing:
\begin{equation*}
\int_0^T \int_{\Omega} u \cdot v\dx \dt
\end{equation*} 
 is defined by: 
\begin{lstlisting}[language=Python,numbers=none]
J = Functional(inner(u, v)*dx*dt)
\end{lstlisting}
Alternatively, the functional can be evaluated at a specific time. For example  
\begin{equation*}
 \int_\Omega u(t=1) \cdot v(t=1) \dx,
\end{equation*}
is implemented by: 
\begin{lstlisting}[language=Python,numbers=none]
J = Functional(inner(u, v)*dx*dt[1])
\end{lstlisting}
Functionals may consist of forms integrated over time, forms evaluated at a particular time, and sums of these. 
This allows the construction of complex objective functionals, e.g. a data assimilation problem with a regularisation term involving the initial condition might use the following objective functional:
\begin{equation*}
 %\int_{\Omega} \left< a(t=T), b(t=T) \right> \dx + \frac{1}{2}\int_{\Omega} f\frac{e - f}{\left< \nabla g, \nabla h \right>}\dx \dt
 \int_0^T \left\|u - u_{\textrm{obs}}\right\|_{L^2(\Omega)}^2 \dt + | u(t=0) |_{H^1(\Omega)}^2.
\end{equation*}
The implementation of that functional is:
\begin{lstlisting}[language=Python,numbers=none]
  J = Functional(inner(u - u_obs, u - u_obs)*dx*dt 
                + inner(grad(u), grad(u))*dx*dt[0]) 
\end{lstlisting}
Similarly, the user specifies the optimisation parameter $m$. 
For example, the following code defines the optimisation parameter to be the initial value of $u$: 
\begin{lstlisting}[language=Python,numbers=none]
      m = InitialConditionParameter(u)
\end{lstlisting}
If a scalar value $s$ is to be optimised, one may use
\begin{lstlisting}[language=Python,numbers=none]
      m = ScalarParameter(s)
\end{lstlisting}

In the final step, the user defines the reduced functional $\hat J$ with
\begin{lstlisting}[language=Python,numbers=none]
      J_hat = ReducedFunctional(J, m)
\end{lstlisting}
In order to optimise for multiple parameters simultaneously, the user can optionally pass a list of \texttt{Parameter} objects.

At this point, the reduced functional object \texttt{J\_hat} can be used to solve the associated minimisation problem $\min_m \hat J(m)$ by calling:
\begin{lstlisting}[language=Python,numbers=none]
      m_opt = minimize(J_hat)
\end{lstlisting}
or the associated maximisation problem $\max_m \hat J(m)$ by calling:
\begin{lstlisting}[language=Python,numbers=none]
      m_opt = maximize(J_hat)
\end{lstlisting}
Both of these functions solve the optimisation problem with the default settings and return the optimised parameters when finished. 
Additional arguments may be used to set and configure the optimisation algorithm and to define box, inequality or equality constraints, e.g.: 
\begin{lstlisting}[language=Python,numbers=none]
m_opt = maximize(J_hat, bounds = (u_lower, u_upper), method='SLSQP')
\end{lstlisting}

Currently, the framework supports most of the algorithms in the optimisation package of SciPy~\cite{scipy}.
For problems without additional constraints these are: 
\begin{itemize}
\item the Nelder-Mead method~\cite{nelder1965};
\item a modification to Powell's method~\cite{powell1964};
\item a nonlinear conjugate gradient method~\cite[\S 5.2]{wright2006};
\item the BFGS method~\cite[\S 6.1]{wright2006}; 
\item the Newton-CG method~\cite[\S 7.1]{wright2006} 
\item and simulated annealing~\cite{laarhoven1987}.
\end{itemize}
Both Powell's method and simulated annealing are gradient-free optimisation algorithms. 
For problems with box, inequality or equality constraints, the user has the choice between:
\begin{itemize}
 \item sequential quadratic programming~\cite{kraft1994};
\item the gradient-free Constrained Optimization by Linear Approximation method~\cite{powell1994}.
\end{itemize}
Finally, if only box-constraints are present, the 
\begin{itemize}
 \item L-BFGS-B method~\cite[\S 7.2]{wright2006};
 \item a Newton-CG implementation that supports box constraints~\cite{nash1984},
\end{itemize}
 may be used. 
 
The user also has access to more advanced features, such as to automatically verify each gradient computation with the Taylor remainder convergence test
during the optimisation procedure, or 
to execute user-supplied code after each gradient computation, for example to create convergence plots.

\subsubsection{Implementation}
%Internally, it as an interface between the forward model implemented in FEniCS, its adjoint model that is automatically derived by \libadjoint and
%the optimisation algorithm. 
%The optimisation framework presented here is implemented as part of \da and heavily relies on \libadjoint's annotation and callbacks. 

The \texttt{ReducedFunctional} class provides two main functionalities: 
the evaluation of the objective functional for a given parameter value, and the computation of the functional gradient. 

The functional evaluation is implemented by solving the forward equations that have been stored on \libadjoint's tape and simultaneously computing the objective functional. 
However, the tape contains the parameter values of the initial forward run and a na\"{i}ve replay would reevaluate the 
forward model with the original parameters. 
This issue is resolved by first modifying the tape so that it reflects the most recent parameter values before executing the functional evaluation.

The implementation of the gradient computation relies on the adjoint derivation of \libadjoint and \da to compute the gradient with the adjoint approach. 
Optionally, the user may use a checkpointing scheme to balance the storage and recomputation cost of the gradient computation~\cite{griewank1992,farrell2012}.

The \texttt{minimize} and \texttt{maximize} routines implement the interface to the optimisation algorithms. 
Optimisation methods typically require the implementation of the functional evaluation and its gradient. 
The \texttt{minimize} and \texttt{maximize} routines generate these functions using the \texttt{ReducedFunctional} object.
This involves the conversion of \dolfin data structures to generic array data types on which the optimisation algorithms operate. 

%In PDE-constrained optimisation, each functional and gradient evaluation involves the solution of one or more partial differential equations.
Parallel execution is often crucial in PDE-constrained optimisation to achieve reasonable run times. 
While \dolfin supports the parallel solution of the forward and adjoint models, the considered implementations of the optimisation algorithms are not designed for distributed execution.
The \texttt{minimize} and \texttt{maximize} routines circumvent this problem by executing the PDE solves in parallel, 
but replicating the optimisation computation on each processor. %running the optimisation algorithm in serial. 
For that, the \texttt{minimize} and \texttt{maximize} functions serialise the distributed data structures of the optimisation parameters, 
the functional gradient and the constraints, as these are used by the optimisation algorithm.
All other variables, in particular the forward and adjoint solutions, remain distributed and are not communicated. 
%This approach requires serialise the functional gradient after every gradient computation.
This approach works well for small-scale optimisation problems, where the communication time for 
gathering the data and the execution time spent in the 
optimisation algorithm is small compared to the runtime of the PDE solves.
For large-scale optimisation problems one should consider interfacing a 
parallel optimisation algorithm, such as TAO~\cite{tao2012} or OPT++~\cite{meza1994}.

Finally, there are cases where optimisation based purely on \libadjoint's tape is not desired. 
For example, a forward model with an adaptive timestepping scheme changes the timestep according to certain conditions.
This adaptivity is not reflected in the tape, and hence the optimisation process would only use the timestep choice that was used to build the tape.
In such cases, the user can manually implement the functional evaluation; however, this approach has the disadvantage that 
the computed gradients become inconsistent with the discrete forward model, which can result in a reduced convergence of the optimisation algorithm.

\subsection{Restrictions}\label{sec:restrictions}
The first restrictions are those associated with the adjoint computation \cite[\S5.4]{farrell2012}.
In particular, the automated adjoint derivation relies on the differentiability of the forward model,
and that the forward model is implemented entirely in the Python interface to DOLFIN. 

State constraints other than the PDE constraint are not handled
automatically. However, the framework does provide many of the tools required to implement the solution
of such problems. An example is presented in section \ref{sec:mpec}, where a problem with a variational
inequality is broken down into a sequence of PDE-constrained optimisation problems, each of which is solved
with the framework presented here.

Another restriction is that shape optimisation is not yet automated. It is possible to apply the shape
calculus approach \cite{schmidt2010,schmidt2011b} which derives an expression for the shape gradient
of the functional in terms of the forward and adjoint solutions, which may then be computed using the
automatically generated adjoint model. In future work, we plan to automate this shape calculus analysis
to extend the framework to support automated shape optimisation.

\section{Example Code}\label{sec:examples}
This section demonstrates the application of the presented optimisation framework to two optimal control problems.
The first example solves the optimal heating problem~\eqref{eq:optimal_control_with_heat_equation_con_example}. 
The governing PDE in this case is the stationary heat equation.
The second example replaces the stationary heat equation with a time-dependent, nonlinear PDE. 
Although this problem adds significant complexity to the forward and adjoint PDEs, 
the required code changes are minimal.

\subsection{Distributed control of the heat equation}\label{distributed_control_of_the_heat_equation}
% Hinze page 56

The first example solves the optimal control problem of the stationary heat equation~\eqref{eq:optimal_control_with_heat_equation_con_example}.
The problem possesses a unique optimal solution if the box constraints $a$ and $b$ are bounded~\cite[chapter 2.5.1]{troeltzsch2005}.
Otherwise the regularisation term must be strictly positive, i.e. $\alpha > 0$, to ensure uniqueness.
In the following example, these conditions are satisfied by choosing $\alpha = 0$, $a = 0$ and $b = 0.5$.

To begin with, the user implements and solves the forward problem.
By doing so, the optimisation framework creates the tape of the forward model.
This tape is used by the optimisation framework to compute all future functional evaluations and gradient computations. 

\begin{figure}[t]
\centering
        \subfloat[Desired temperature profile $u_d$]{
                \centering
                \includegraphics[width=0.4\textwidth]{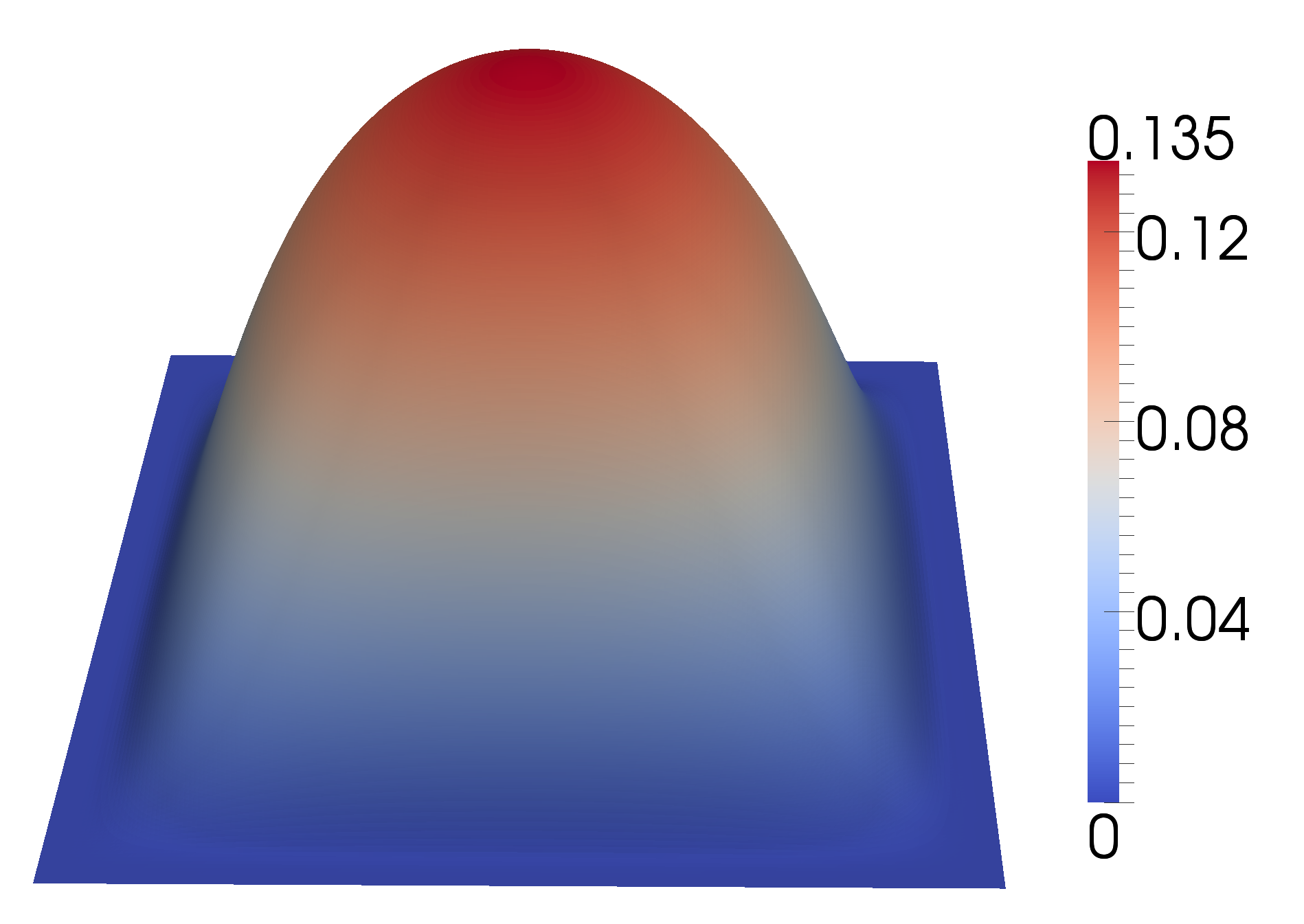}
                \label{fig:optimal_control/bump_function/desired_state_paraview.png}
        }
        \subfloat[Optimised temperature profile $u$]{
                \centering
                \includegraphics[width=0.39\textwidth]{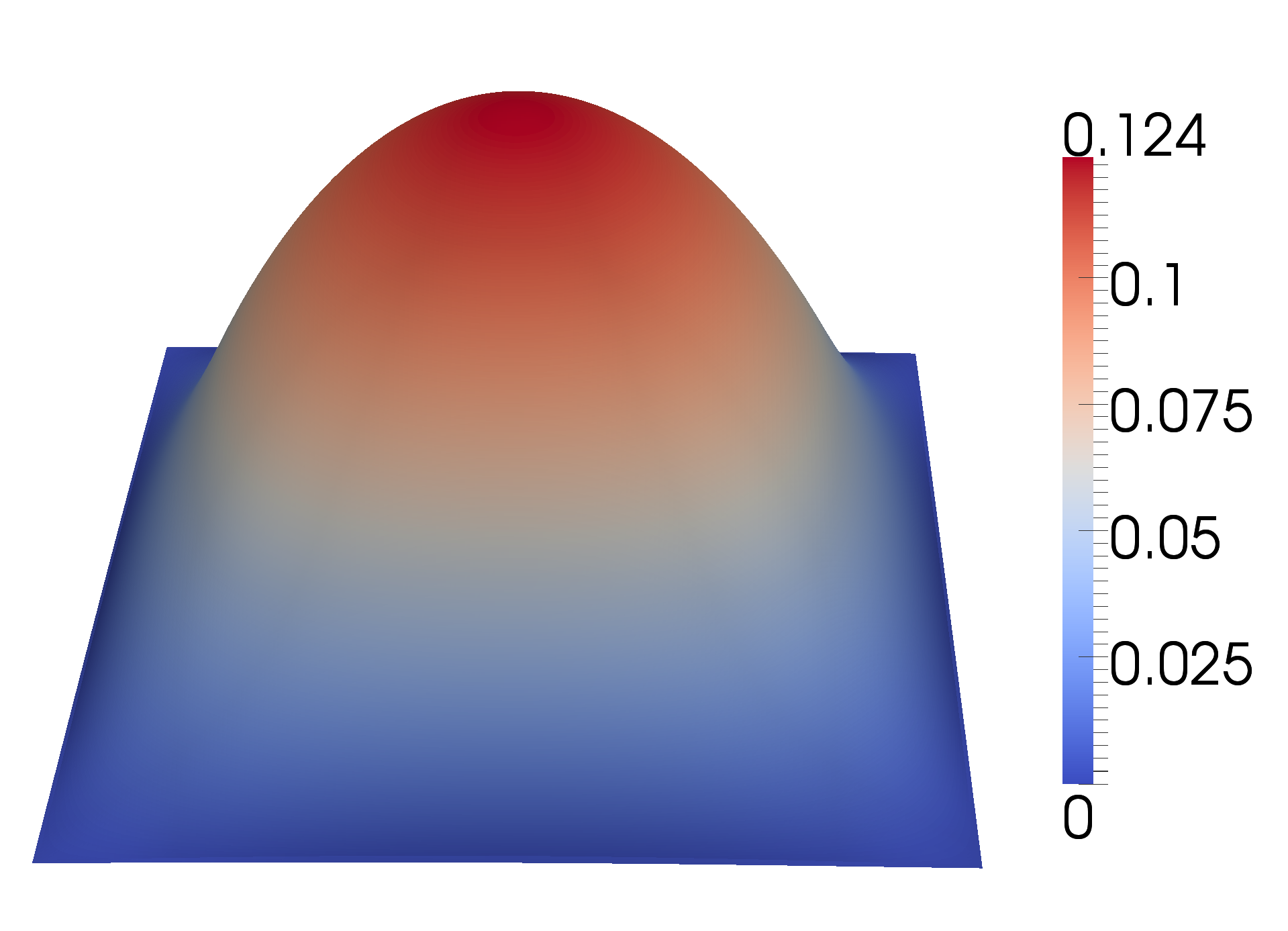}
                \label{fig:optimal_control/bump_function/optimal_state_paraview.png}
        }
        \\
        \subfloat[Difference between optimised and desired temperature profiles $u-u_d$]{
                \centering
                \includegraphics[width=0.4\textwidth]{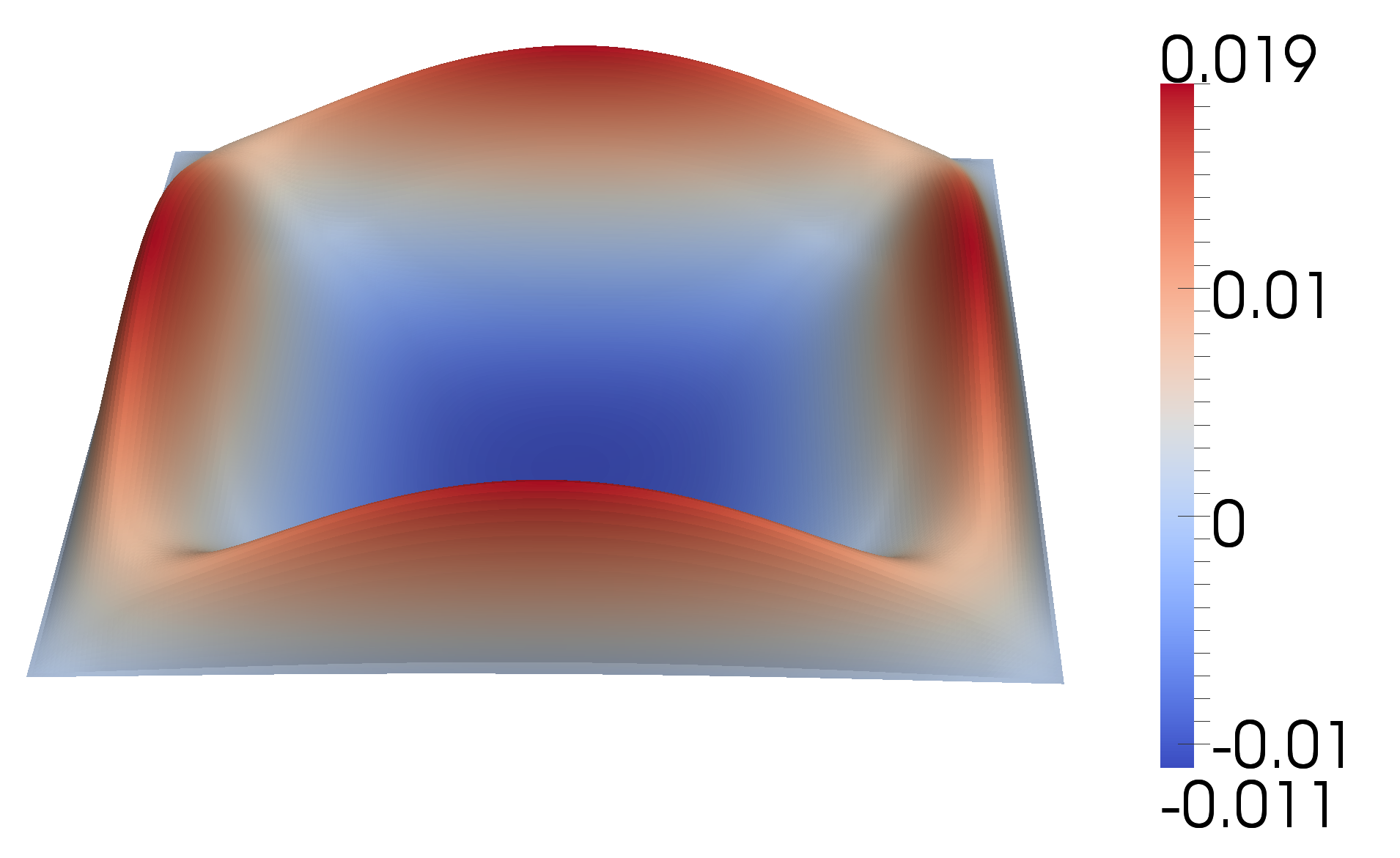}
                \label{fig:optimal_control/bump_function/difference_between_optimal_and_desired_state_paraview.png}
        }
        \subfloat[Optimised heat source control $m$]{
                \centering
                \includegraphics[width=0.4\textwidth]{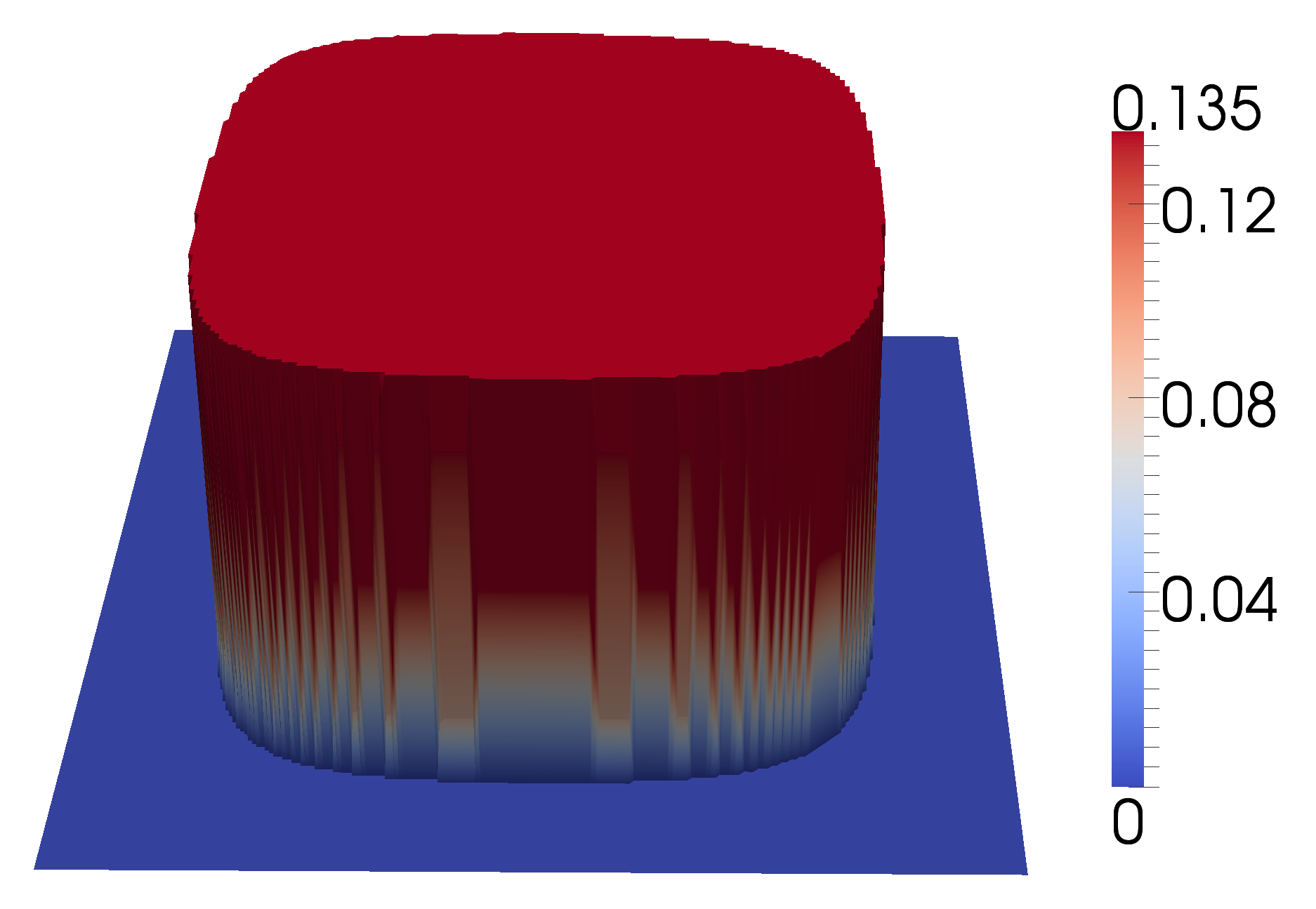}
                \label{fig:optimal_control/bump_function/optimal_control_paraview.png}
        }
        \caption{The solutions of the stationary optimal heating problem with box constraints. 
                 The heat source control is limited by the box constraints in large parts of the domain,
                  which leads to a relatively large difference between the desired and optimised temperature profiles.}\label{fig:solution_optimal_heating_problem}
\end{figure}

For this example the two-dimensional domain $\Omega \equiv [-1, 1]^2$ was uniformly discretised using triangular elements. 
The finite-dimensional function spaces are constructed using \ponecg elements for the PDE solution $u$ and the desired temperature profile $u_d$, and \pzerodg elements for the heat source $m$.
The latter choice is motivated by the fact that the control can possess discontinuities for $\alpha = 0$.
The following Python code initialises the domain, the function spaces and all required functions in \dolfin:

\begin{lstlisting}[language=Python,linerange={9-18}]
# Define the domain [-1, 1] x [-1, 1]
n = 200
mesh = RectangleMesh(-1, -1, 1, 1, n, n)
# Define the function spaces
V = FunctionSpace(mesh, "CG", degree = 1)
W = FunctionSpace(mesh, "DG", degree = 0)
# Define the functions
u = Function(V, name = "Solution")
m = Function(W, name = "Control")
v = TestFunction(V)
\end{lstlisting}
The weak form of the heat equation is obtained by multiplying the PDE with a test function $v \in V$, then integrating the result over the domain and integrating by parts.
The resulting weak formulation is: find $u \in V$ such that: 
\begin{equation*}
\left<\nabla u, \nabla v\right>_\Omega = \left<m, v\right>_\Omega \quad \forall v \in V. 
\end{equation*}
The associated code resembles the mathematical notation closely:
\begin{lstlisting}[language=Python,linerange={46-50},firstnumber=11]
# Solve the forward model to create the tape 
F = (inner(grad(u), grad(v)) - m*v)*dx 
bc = DirichletBC(V, 0.0, "on_boundary")
solve(F == 0, u, bc)
\end{lstlisting}

Next, the objective functional is defined. 
In this example, the desired temperature profile $u_d$ is defined as:
\begin{equation*}
    u_d(x, y) = e^{-1/(1-x^2) - 1/(1-y^2)}, 
\end{equation*}
plotted in figure~\ref{fig:optimal_control/bump_function/desired_state_paraview.png}. 
The relevant code is:
\begin{lstlisting}[language=Python,linerange={52-54},firstnumber=15]
# Define the functional of interest
u_d = exp(-1/(1-x[0]*x[0])-1/(1-x[1]*x[1]))
J = Functional((0.5*inner(u-u_d, u-u_d))*dx)
\end{lstlisting}

At this point, the optimal control problem can be solved: 
\begin{lstlisting}[language=Python,linerange={58-63},firstnumber=18]
# Define the reduced functional
J_hat = ReducedFunctional(J, SteadyParameter(m))
# Solve the optimisation problem
m_opt = minimize(J_hat, method = "L-BFGS-B", bounds = (0.0, 0.5), 
                 options = {"gtol": 1e-16, "ftol": 1e-16})
\end{lstlisting}

The last parameter of \texttt{minimize} sets the termination condition 
to stop if either the infinity norm of the projected gradient or the relative change of the functional value 
drops below the specified tolerance.

% Reproduce numerical results by running the python script in 'optimal_control' subfolder in this directory. 

The optimisation algorithm starts with a zero estimate for the control, i.e. $m^{(0)} = 0$.
The corresponding functional value is $\hat J\left(m^{(0)}\right) = 8.9 \times 10^{-3}$. 
The results of the optimisation are shown in figure \ref{fig:solution_optimal_heating_problem}. 
The final value of the objective functional is $1.4 \times 10^{-4}$. 
The desired and optimised temperature profile are of similar shape, but their maximum values differ significantly. 
This is reflected in the fact that the box constraints are active in large parts of the domain (figure~\ref{fig:optimal_control/bump_function/optimal_control_paraview.png}).

\begin{figure}[t]
\centering
        \subfloat[Desired temperature profile $u_d$]{
                \centering
                \includegraphics[width=0.4\textwidth]{optimal_control/bump_function/desired_state_paraview.png}
        }
        \subfloat[Optimised temperature profile $u$]{
                \centering
                \includegraphics[width=0.4\textwidth]{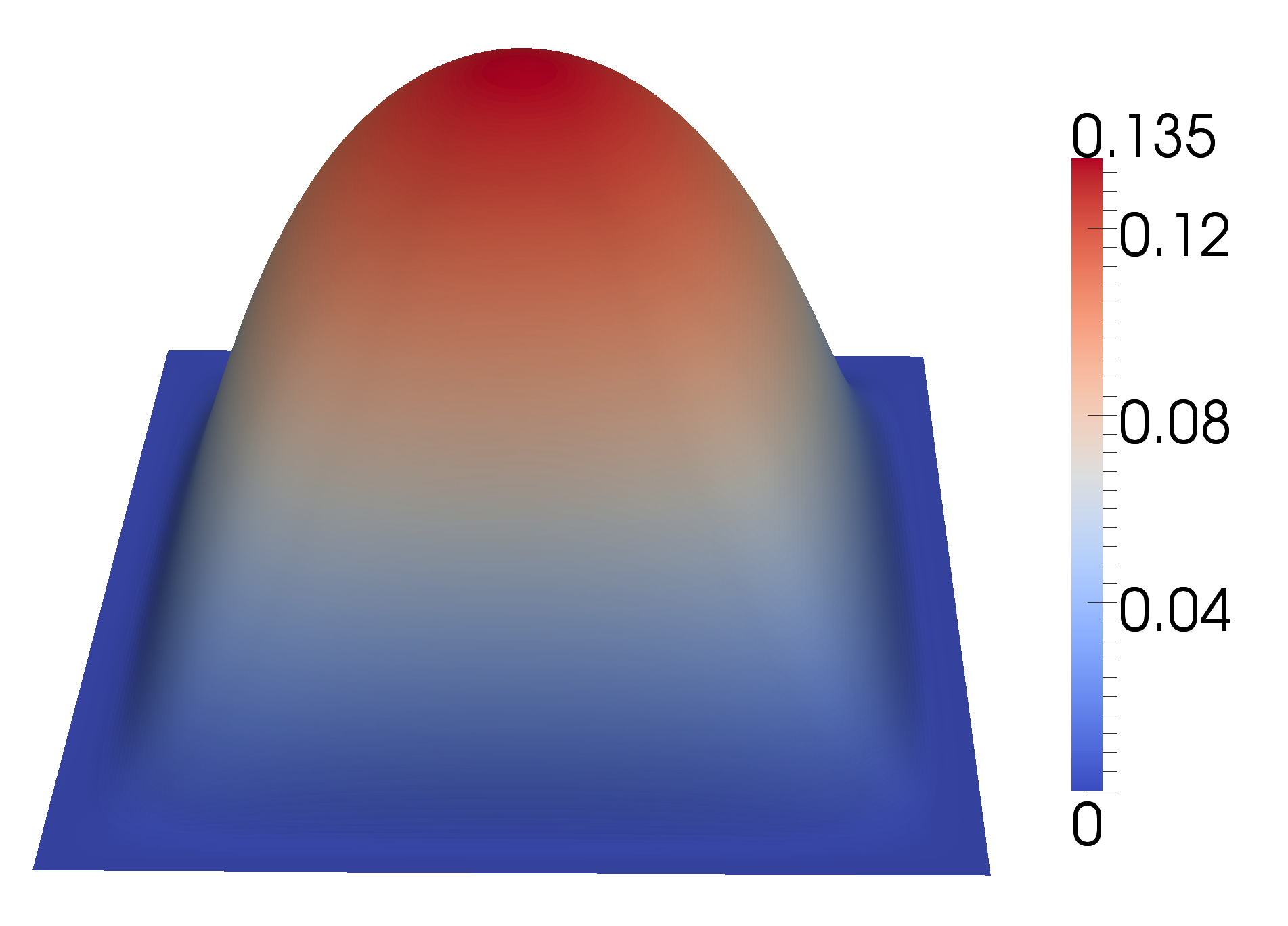}
                \label{fig:optimal_control/bump_function/optimal_state_without_bounds_paraview.png}
        }
        \\
        \subfloat[Difference between optimal and desired temperature profiles $u-u_d$]{
                \centering
                \includegraphics[width=0.4\textwidth]{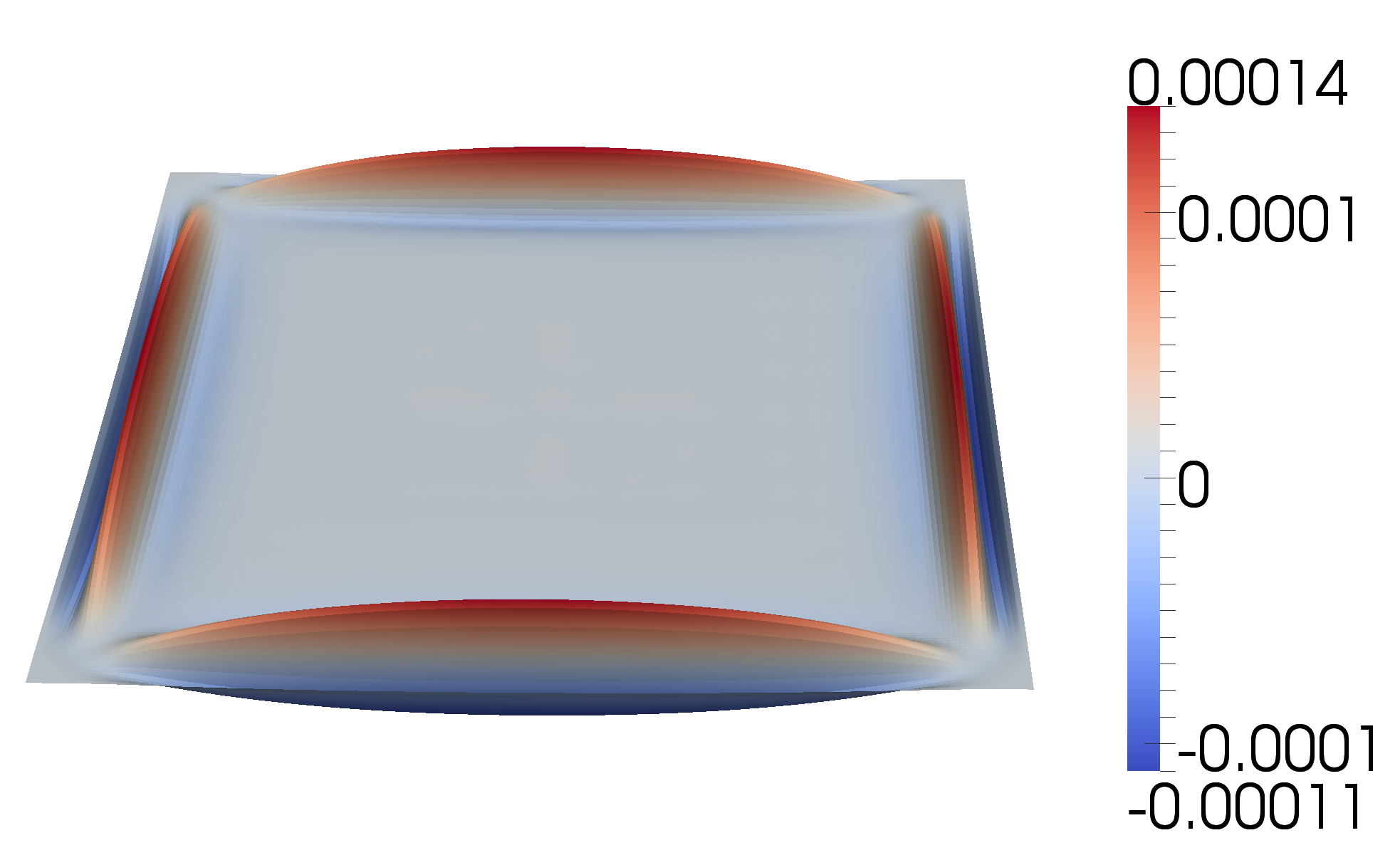}
                \label{fig:optimal_control/bump_function/difference_between_optimal_and_desired_state_without_bounds_paraview.png}
        }
        \subfloat[Optimised heat source control $m$]{
                \centering
                \includegraphics[width=0.4\textwidth]{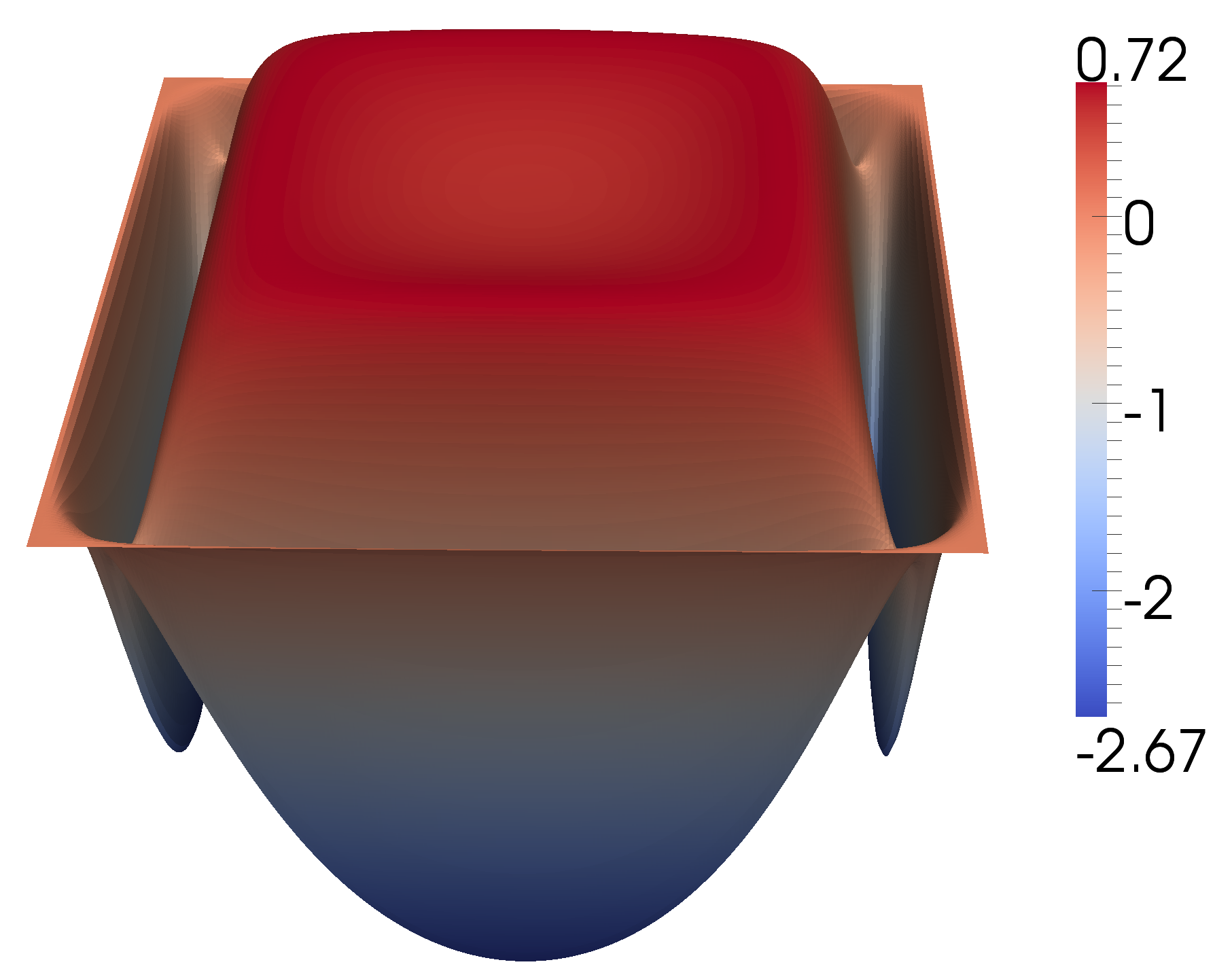}
                \label{fig:optimal_control/bump_function/optimal_control_without bounds_paraview.png}
        }
        \caption{The solutions of the stationary optimal heating problem without box constraints. The optimised heat source control achieves a good agreement between the desired and optimised temperature profiles.}\label{fig:solution_optimal_heating_problem_without_bounds}
\end{figure}

By removing the box constraints, a better agreement between the desired and optimised temperature profiles is expected.
Therefore the box constraints were removed and the regularisation parameter set to $\alpha = 10^{-7}$, in order to ensure the uniqueness of the optimal solution.
The results are shown in figure \ref{fig:solution_optimal_heating_problem_without_bounds}. 
Compared to the previous results, the pointwise difference between the desired and optimised temperature profiles is significantly decreased,
yielding a final functional is $1.6 \times 10^{-7}$.

\subsection{Distributed control of a nonlinear, time-dependent PDE}
% Hinze page 56

This example modifies the optimal heating problem by replacing the stationary heat equation with a time-dependent, nonlinear PDE:
\begin{equation}
\begin{aligned}
\min_m~&\frac{1}{2} \| u(t=T) - u_d \|^2_{L^2(\Omega)} + \frac{\alpha}{2} \| m \|^2_{L^2({\Omega})} \\ 
 \textrm{subject to } 
 & \frac{\partial u}{\partial t} -\nabla^2 u + u^3 = m \hspace{0.5cm} \textrm{ on } \Omega \times (0, T], \\ 
 & u = 0 \hspace{2.8cm} \textrm{ on } \partial \Omega \times (0, T],  \\
 & u = 0 \hspace{2.8cm} \textrm{ for } \Omega \times \{0\},  \\
 & a  \le m \le b \hspace{2.07cm} \textrm{ on } \partial \Omega \times (0, T],
\end{aligned}\label{eq:optimal_control_transient}%
\end{equation}
where $u: \Omega \times (0, T] \to \mathbb R$ and $u_d: \Omega \times (0, T] \to \mathbb R$ are the time-dependent PDE solution and desired temperature profile, respectively, 
and the control $m: \Omega \to \mathbb R$ is constant in time.
The nonlinearity and time-dependency of the new governing PDE
adds significant complexity to the solution process of the optimisation algorithm.
In particular, the gradient computations involve the storage of the forward solution trajectory (or the application of some checkpointing scheme) and 
the solution of the associated time-dependent adjoint system. 

However, since the steps for recording the tape and the functional and gradient evaluations are automated,
the code from the previous example can be reused almost entirely to solve problem~\eqref{eq:optimal_control_transient}. 
The only modification required is to replace the weak formulation of the heat equation with the new governing PDE. 
The following Python code shows an example implementation with a backward Euler time discretisation and a Newton solver for the nonlinear equation solve at each timestep:
\begin{lstlisting}[language=Python,linerange={35-42},firstnumber=11]
# Define the weak form
timestep = 0.01
F = ((u - u_old)*v/timestep + inner(grad(u), grad(v)) + u**3*v - m*v)*dx 
# Perform 10 timesteps
for t in range(11):
    solve(F == 0, u, bc)
    u_old.assign(u)
    adj_inc_timestep()
\end{lstlisting}

The optimisation framework can also use a checkpointing scheme to reduce the storage cost of the gradient computations.
For example, the multistage checkpointing scheme developed by~\citeN{stumm2009}, which supports storing checkpoints both in RAM and on disk, is activated with:
\begin{lstlisting}[language=Python,numbers=none]
adj_checkpointing(strategy='multistage', steps=11, 
                    snaps_on_disk=2, snaps_in_ram=3)
\end{lstlisting}

\section{Verification}\label{sec:verification}
This section applies the optimisation framework to problems with analytical solutions in order to compare the numerical order of convergence with the theoretical expectation.
An agreement of the convergence rates is considered to be a strong  indicator that the implementation is correct \cite{salari2000}.

The considered analytical solutions are based on the optimal control problem \eqref{eq:optimal_control_with_heat_equation_con_example} of the heat equation, extended with an additional source term $s: \Omega \to \mathbb R$:
\begin{equation}
\begin{aligned}
\min_m~&\|u -u_d\|_{L^2(\Omega)}^2 \\ 
 \textrm{subject to } & 
 - \nabla^2 u = m + s && \textrm{in } \Omega, \\%\label{eq:optimal_control_mms_state_equation} \\
                      & u  = 0 && \textrm{on } \partial \Omega, \\
                      & -1  \le m \le 1 && \textrm{in } \Omega. &&
\end{aligned}\label{eq:optimal_control_mms_test}%
\end{equation}
%that the conditions for uniqueness hold, see section \ref{distributed_control_of_the_heat_equation}.

The following sections perform the two convergence tests, the first one with a continuous optimal control function, and the second one with a discontinuous optimal control function.

\subsection{Smooth control}

%Figure \ref{fig:optimal_control_smooth_control} shows a plot of this optimal control $m^{\mathrm{opt}}$.
%\begin{figure}[t]
%        \centering
%        \includegraphics[width=0.5\textwidth]{optimal_control/optimal_control_mms/analytical_control.png}
%        \caption{The optimal control $m^{\mathrm{opt}}$ for the smooth control test.}
%        \label{fig:optimal_control_smooth_control}
%\end{figure}
% Mesh element size: [0.125, 0.0625, 0.03125, 0.015625, 0.0078125]
%Control errors: [0.09536297528876767, 0.03695161851170893, 0.017090624580624107, 0.00832949849840525, 0.004108343007252608]
%Control convergence: [1.3677916910405856, 1.1124324354172017, 1.0369035799016157, 1.0196729986811346]
%State errors: [0.0003825166740543751, 9.032966495746064e-05, 2.2049544176352937e-05, 5.432907991172374e-06, 1.3569318892689531e-06]
%State convergence: [2.0822508740344885, 2.0344510259940574, 2.020952312070796, 2.0013763075124604]
%Generating convergence plots

\begin{table}[t]
\centering
  \begin{tabular}{ccccc}
    Element size & \small{$\left\| m - m^{\mathrm{opt}}\right\|$} & order & \small{$\left\|u-u^{\mathrm{opt}} \right\|$} & order \\
    \hline
    $1.25 \times 10^{-1}$ & $9.54 \times 10^{-2}$ &  & $3.83 \times 10^{-4}$ & \\
    $6.25 \times 10^{-2}$ & $3.70 \times 10^{-2}$ & $1.34$ & $9.03 \times 10^{-5}$ & $2.08$ \\
    $3.12 \times 10^{-2}$ & $1.71 \times 10^{-2}$ & $1.11$ & $2.20 \times 10^{-5}$ & $2.03$ \\
    $1.56 \times 10^{-2}$ & $8.33 \times 10^{-3}$ & $1.04$ & $5.43 \times 10^{-6}$ & $2.02$ \\
    $7.81 \times 10^{-3}$ & $4.11 \times 10^{-3}$ & $1.02$ & $1.36 \times 10^{-6}$ & $2.00$ \\
  \end{tabular}
\caption{The rate of convergence for the smooth control test. The control error shows the expected first order convergence,
and the PDE solution converges as expected at second order.}
  \label{fig:rate_of_convergence_smooth}
\end{table}

The first test is based on an analytical solution with a smooth optimal control function:
\begin{equation*}
 \begin{aligned}
  m^{\mathrm{opt}}(x, y) & \equiv \sin\left(\pi x\right) \sin\left(\pi y\right), \\
u^{\mathrm{opt}}(x, y) & \equiv u_d^{\mathrm{opt}}(x, y) \equiv \frac{1}{2\pi^2}\sin(\pi x) \sin(\pi y), \\
                       s & \equiv 0.
 \end{aligned}
\end{equation*}
It is easy to see that this choice forms an optimal solution to problem~\eqref{eq:optimal_control_mms_test}.

The discretisation and optimisation parameters are configured identically to the setup described in section \ref{distributed_control_of_the_heat_equation}. 
Then the convergence test was performed on five uniformly discretised meshes with decreasing mesh element sizes.
The resulting errors and convergence rates are given in table~\ref{fig:rate_of_convergence_smooth}.
The first-order convergence for the control solution $m$ is expected as the underlying function space is discretised with \pzerodg finite elements. 
Similarly, a second-order rate of convergence is observed  for the PDE solution $u$, as it is discretised with \ponecg finite elements.

\subsection{Bang-bang control}
The second verification test is motivated by the fact that box constraints can lead to optimal control solutions with discontinuities.
The following test, derived in \citeN[chapter 2.9.1]{troeltzsch2005}, yields an optimal control function with discontinuities in a chessboard-like shape:
%\begin{figure}
%        \begin{subfigure}[b]{0.5\textwidth}
%                \centering
%                \includegraphics[width=\textwidth]{optimal_control/optimal_control_bang_bang_mms/rate_of_convergence_for_the_control_solution.png}
%                \caption{Rate of convergence for the control solution.}
%                \label{fig:rate_of_convergence_bang_bang_for_the_control_solution}
%        \end{subfigure}
%        ~
%        \begin{subfigure}[b]{0.5\textwidth}
%                \centering
%                \includegraphics[width=\textwidth]{optimal_control/optimal_control_bang_bang_mms/rate_of_convergence_for_the_state_solution.png}
%                \caption{Rate of convergence for the state solution}
%                \label{fig:rate_of_convergence_bang_bang_for_the_state_solution}
%        \end{subfigure}
%        \caption{Convergence plot for the bang bang control test.}\label{fig:rate_of_convergence_bang_bang}
%\end{figure}
\begin{equation*}
m^{\mathrm{opt}}(x, y) \equiv - \mathrm{sign} \left(-\sin\left(8\pi x\right)\sin\left(8\pi y\right)\right).
\end{equation*}
This kind of control, where the control values jump from one box constraint limit to the other, is also known as bang-bang control.
The optimal state solution is chosen to be:
\begin{equation*}
u^{\mathrm{opt}}(x, y) \equiv \sin(\pi x) \sin(\pi y).
\end{equation*}
By applying the optimality conditions, the source term $s$ and the desired PDE solution $u_d$ are obtained:
\begin{equation*}
s(x, y) \equiv 2 \pi^2 \sin(\pi x)\sin(\pi y) + \mathrm{sign}\left(-\sin\left(8\pi x\right)\sin\left(8\pi y\right)\right),
\end{equation*}
and:
\begin{equation*}
u_d(x, y) \equiv \sin(\pi x)\sin(\pi y) + \mathrm{sign}\left(-\sin\left(8\pi x\right)\sin\left(8\pi y\right)\right).
\end{equation*}

The convergence test is performed with the same configuration as in the previous test. 
The resulting errors and convergence rates are given in table~\ref{fig:rate_of_convergence_bang_bang}.
It shows the expected first-order convergence for the control and second-order convergence for the state solution,
indicating that the optimisation implementation is correct.

%\begin{figure}[t]
%        \centering
%        \includegraphics[width=0.5\textwidth]{optimal_control/optimal_control_bang_bang_mms/analytical_control.png}
%        \caption{The optimal control $m^{\mathrm{opt}}$ for the bang-bang test.}%
%        \label{fig:optimal_control_bang_bang_control}%
%\end{figure}
%i = 5
%element size: 0.03125
%Control errors: [0.47642203625438884]
%State errors: [0.0007376927839030025]
%
%control convergence: 0.9863614698652683
%u convergence: 1.8786403813092465
%
%i = 6
%element size: 0.015625 
%Control errors: [0.2404736260029515]
%State errors: [0.00020060808370866127]
%
%control convergence: 0.9995789554436296
%u convergence :1.9932932989076237
%
%i = 7
%element size: 0.0078125
%Control errors: [0.1202719087370466]
%State errors: [5.038570693004131e-05]
%
%control convergence: 1.182081153033993 
%u convergence : 2.025288582354569
%
%i = 8
%i element size: 0.00390625
%Control errors: [0.05300566884709941]
%State errors: [1.2377551530940244e-05]
%
%control convergence: 1.204376416158396 
%u convergence: 2.285986586450612
%
%i = 9
%Element sizes: [0.001953125]
%Control errors: [0.023002174386830877]
%State errors: [2.537956714859826e-06]
\begin{table}
\centering
  \begin{tabular}{ccccc}
    Element size & \small{$\left\| m - m^{\mathrm{opt}}\right\|$} & order & \small{$\left\|u-u^{\mathrm{opt}} \right\|$} & order \\
    \hline
    $3.13 \times 10^{-2}$ & $4.76 \times 10^{-2}$ &  & $7.38 \times 10^{-4}$ & \\
    $1.56 \times 10^{-2}$ & $2.41 \times 10^{-2}$ & $0.99$ & $2.01 \times 10^{-4}$ & $1.89$ \\
    $7.81 \times 10^{-3}$ & $1.20 \times 10^{-2}$ & $1.00$ & $5.04 \times 10^{-5}$ & $1.99$ \\
    $3.91 \times 10^{-3}$ & $5.30 \times 10^{-3}$ & $1.18$ & $1.24 \times 10^{-5}$ & $2.03$ \\
    $1.95 \times 10^{-3}$ & $2.30 \times 10^{-3}$ & $1.20$ & $2.54 \times 10^{-6}$ & $2.29$ \\
  \end{tabular}
\caption{The rate of convergence for the bang-bang control test. The control error shows the expected first order convergence,
and the PDE solution converges at second order as expected.}
\label{fig:rate_of_convergence_bang_bang}
\end{table}

\section{Applications}\label{sec:applications}

\subsection{Optimal control governed by an elliptic variational inequality} \label{sec:mpec}

In this section we apply the framework to an optimal control problem investigated in \citeN[\S5.2]{hintermueller2011}.
This problem involves a variational inequality constraint on the state, and is an example of a mathematical program
with equilibrium constraints (MPEC). Let $K = \{v \in H^1_0(\Omega): v \ge 0\}$. The problem is stated as:
\begin{subequations}
\begin{align}
\min_{u, m}\ J(u, m) &= \frac{1}{2} \left|\left|u - u_d\right|\right|^2_{L^2(\Omega)} + \frac{\nu}{2} \left|\left|m\right|\right|^2_{L^2(\Omega)} & \\
 \textrm{subject to}\ \  & u \in K, \\
                   & \left<\nabla u, \nabla (v - u)\right>_\Omega \ge \left<f + m, v - u\right>_\Omega \ \forall \ v \in K, \label{eq:mpec_VI} \\
                   & a \le m \le b \ \ \mathrm{a.e.\ in}\ \Omega,
\end{align} \label{eq:mpec_A}%
\end{subequations}
where $u :\Omega \to \mathbb R^2$ is the state variable, $u_d :\Omega \to \mathbb R^2$ is a prescribed state to be matched, $m : \Omega \to \mathbb R$ is the control variable to be determined, $f :\Omega \to \mathbb R$ is a prescribed source term, $a\in \mathbb R$ and $b\in \mathbb R$ are the lower
and upper bounds on the control, and $\nu \in \mathbb R$ is a regularisation parameter. Note that the inequality $u \ge 0\ a.e.$ is implied by $u \in K$.
Following the penalisation approach \cite{tremolieres1981,hintermueller2011}, the variational inequality can be approximated by the penalised equation:
\begin{equation} \label{eq:mpec_B}
(\nabla u, \nabla v) + \frac{1}{\alpha} (-\max{(0, -u)}, v) = (f + m, v) \quad \forall \ v \in H^1_0(\Omega),
\end{equation}
where $\alpha > 0$ is the penalty parameter. It is well known that the solution $u_{\alpha}$ of \eqref{eq:mpec_B} converges to that of the
variational inequality \eqref{eq:mpec_VI} as $\alpha \downarrow 0$. As the $\max$ operator is not differentiable, it is
regularised in turn with a smoothing parameter $\epsilon > 0$ (the ``global'' regularisation of \citeN[equation (2.4)]{hintermueller2011}).
Therefore, to solve \eqref{eq:mpec_A}, a sequence of problems are solved where the variational inequality \eqref{eq:mpec_VI} is replaced
with the regularised penalised equality constraint \eqref{eq:mpec_B}, and the penalisation parameter $\alpha$ is driven to zero. The
solution of one iteration is used as the initial guess for the next.

\begin{figure}
        \centering
        \includegraphics[width=0.75\textwidth]{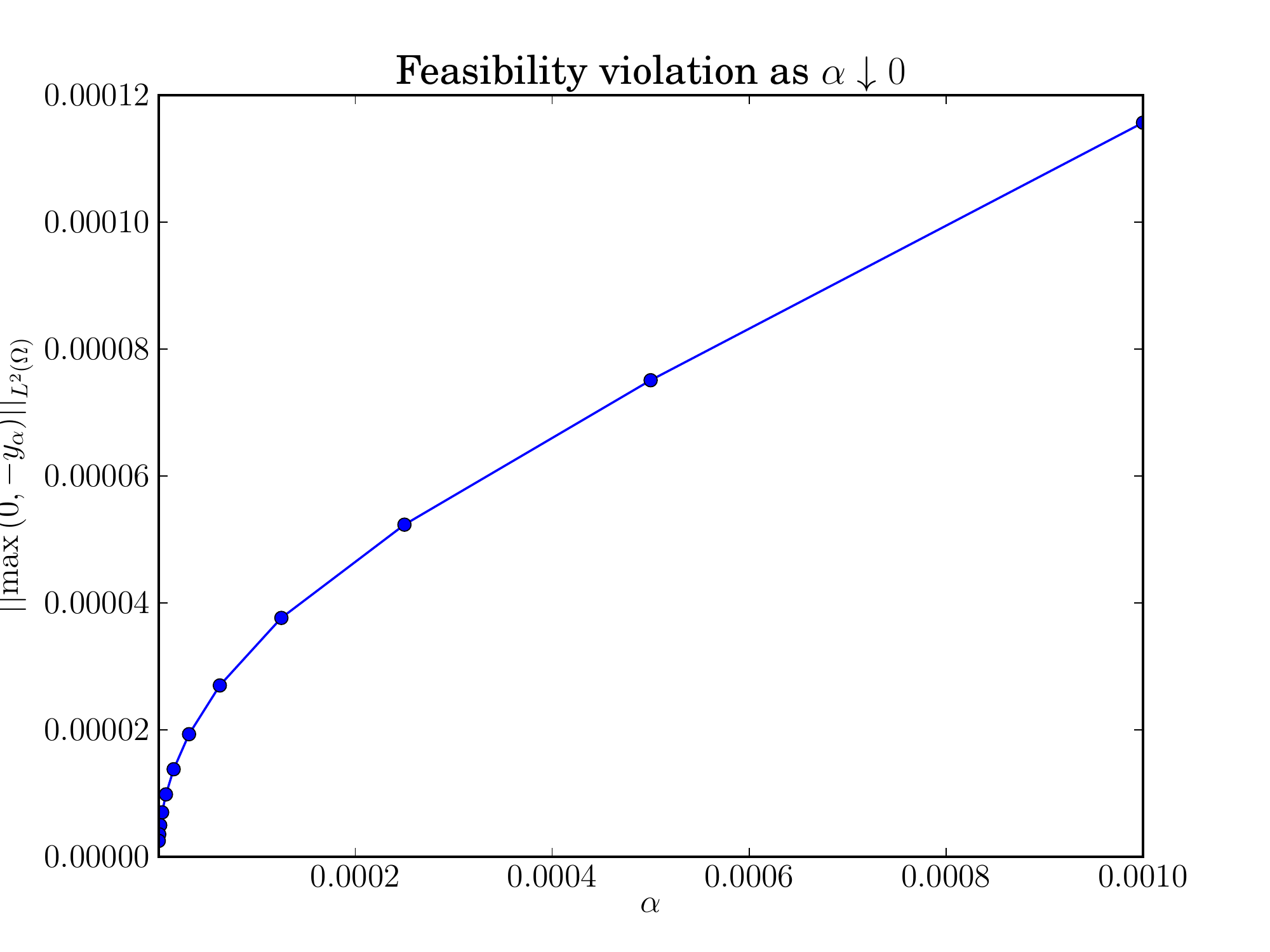}
        \caption{The feasibility violation of the state solution $u$ as a function of $\alpha$. As $\alpha$ approaches 0, the variational inequality
        $u \ge 0\  \mathrm{a.e.\ in}\ \Omega \subset \mathbb R^2$ is enforced.}
        \label{fig:mpec_feasibility}
\end{figure}

The PDE constraint is discretised using linear finite elements for the state and control. The PDE is first solved once with a zero control
$m$ to build a tape of the forward model, and then all subsequent steps are performed by operating on this tape. The value of $\epsilon$ is set to
$10^{-4}$, while $\nu$ is set to $10^{-2}$. The remaining parameters are taken from \citeN[\S 5.2]{hintermueller2011}). The value of $\alpha$ is initialised
to $10^{-3}$ and halved at each penalisation iteration. The penalised subproblem is solved with a call to \texttt{minimize}, which applies
a limited-memory BFGS algorithm with control bounds support \cite{zhu1997b}. The entire program consists of less than 50 lines of code.

\begin{figure}
\centering
        \subfloat[The optimised state $u$]{
                \centering
                \includegraphics[width=0.49\textwidth]{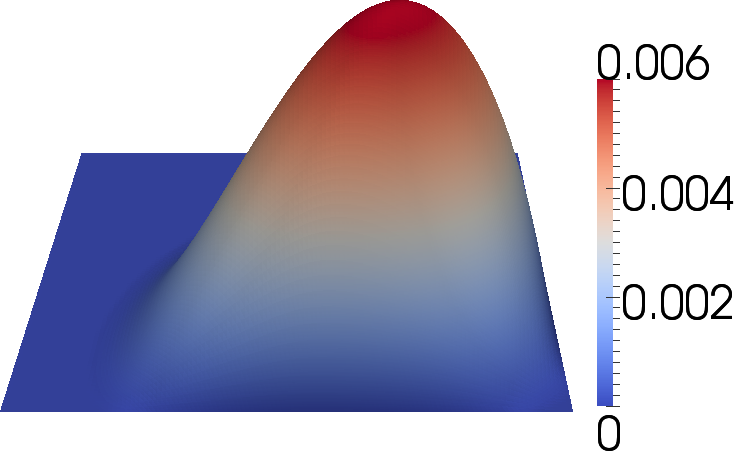}
                \label{fig:mpec-state}
        }
        \subfloat[The optimised control $m$]{
                \centering
                \includegraphics[width=0.49\textwidth]{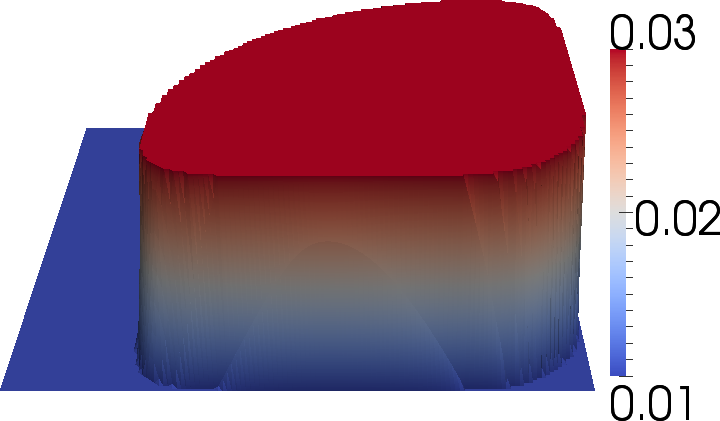}
                \label{fig:mpec-control}
        }
        \caption{The solutions of the optimal control problem with a variational inequality.}
        \label{fig:mpec_solutions}
\end{figure}

The feasibility of the state is measured by computing the diagnostic
$\left|\left| \max{(0, -u)} \right| \right|_{L^2(\Omega)}$; figure \ref{fig:mpec_feasibility} shows its evolution as a function of $\alpha$. 
The control and state solutions of the optimal control problem
are shown in figure \ref{fig:mpec_solutions}. Excellent agreement is found with the solutions of \citeN[figures 5 and 6]{hintermueller2011}, giving confidence
that the solutions are correct.

Finally, the efficiency of the gradient calculation $\nabla \hat{J}$ is benchmarked. For gradient-based optimisation
algorithms to be practical, the computation of the gradient must be affordable. To investigate the performance of the
adjoint-based gradient calculation, one execution of the forward and adjoint models was timed; this was repeated
several times to ensure the results were representative. The results are shown in table~\ref{tab:mpec-timings}. As the forward
model in this case takes eight Newton iterations to converge, the adjoint should take approximately one eighth the cost
of the forward model, for an ideal $R=(\mathrm{forward} + \mathrm{adjoint})/(\mathrm{forward})$ ratio of $1.125$. The observed
ratio is $1.126$, demonstrating the claim in \citeN{farrell2012} that the gradient calculation with dolfin-adjoint 
approaches optimal theoretical efficiency.
\begin{table}
\centering
\begin{tabular}{ccc}
\toprule
       & Runtime (s) & Ratio \\
\midrule
Forward model &  69.08  &     \\
Forward model + adjoint model & 77.85 & 1.126 \\
\bottomrule
\end{tabular}
\caption{Timings for the MPEC gradient calculation. The efficiency of the adjoint approaches
the theoretical ideal value of 1.125.}
\label{tab:mpec-timings}
\end{table}
\subsection{Optimal placement of tidal turbines}\label{sec:turbine_optimisation}

This application investigates an essential problem in the tidal energy industry.
The core idea is to place turbines in the ocean to extract the kinetic energy of the tidal flow and convert it into electricity.
In order to extract an economically useful amount of power, many turbines (possibly hundreds) must be deployed in the tidal stream.
The question is: how should these turbines be placed in relation to each other to maximise the power extracted?
The strong nonlinear interaction between the turbines, the complicated constraints on the configuration (legal site
restrictions, bounds on the gradient of the seafloor), and the sensitive dependence of the power on the configuration make it difficult to
manually identify an optimal configuration.

This problem is formulated as an optimisation problem
constrained by the stationary, nonlinear shallow water equations with appropriate initial and boundary conditions:
\begin{equation}
\begin{aligned}
 \max_{u, m}~&J(u, m) \\
 \mbox{ subject to } & u \cdot \nabla  u - \nu \nabla^2  u  + g \nabla \eta  = \frac{c_b + c_t(m)}{H} \| u\|  u && \mbox{on } \Omega, \\
                    & \nabla \cdot \left(H u\right) = 0 && \mbox{on } \Omega,  \\
\label{eq:shallow_water_equations}
\end{aligned}
\end{equation}
where $\Omega \subset \mathbb R^2$ is the computational domain, the unknowns $u:\Omega \to \mathbb R^2$ and $\eta : \Omega \to \mathbb R$ are the depth-averaged velocity and the free-surface displacement, 
respectively, $H\in \mathbb R$ is the water depth at rest, $g\in \mathbb R$ is the gravitational force, $\nu \in \mathbb R$ is the viscosity coefficient, 
and $c_b \in \mathbb R$ and $c_t$ represent the quadratic bottom friction and the turbine parameterisation, respectively. In a practical
application, problem formulation \eqref{eq:shallow_water_equations} should of course be extended to the time-dependent shallow water equations to account for the
flood and ebb tides.

The functional of interest $J$ is defined to be the power extracted due to the increased friction in the turbine farm~\cite{ben2007,divett2011}:
\begin{equation}
J( u, m) \equiv \frac{1}{2}\int_{\Omega} \rho c_t(m) \|  u \|^3 \textrm{d} \Omega, \label{eq:functional_steady}
\end{equation}
where $\rho$ is the density of water.

\begin{figure}[t]
\centering
        \subfloat[The computational domain; the feasible turbine positions are coloured pink]{
                \centering
                \includegraphics[width=0.49\textwidth]{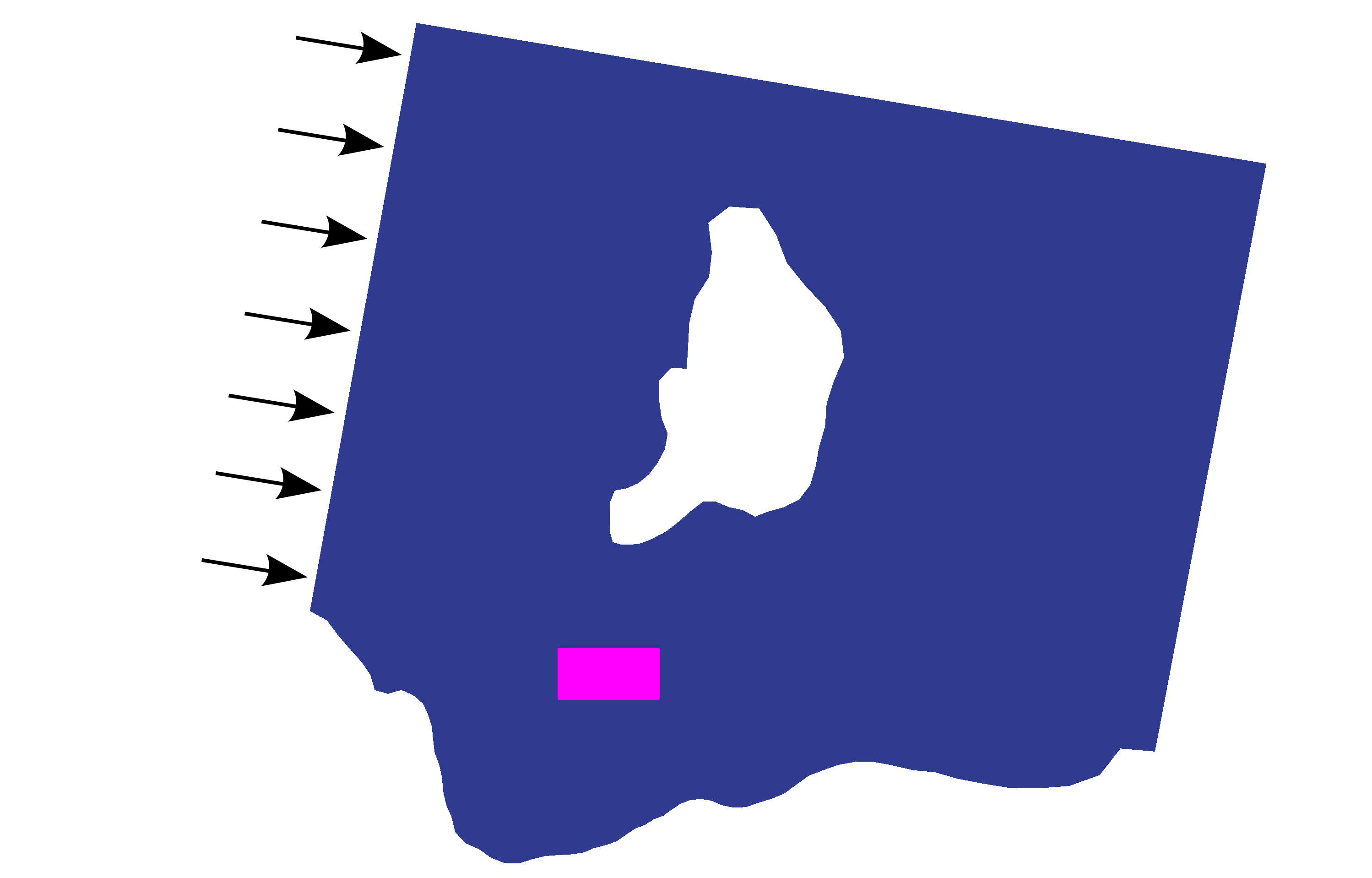}
                \label{fig:orkney_small/domain}
        }
        \subfloat[The objective functional]{
                \centering
                \includegraphics[width=0.49\textwidth]{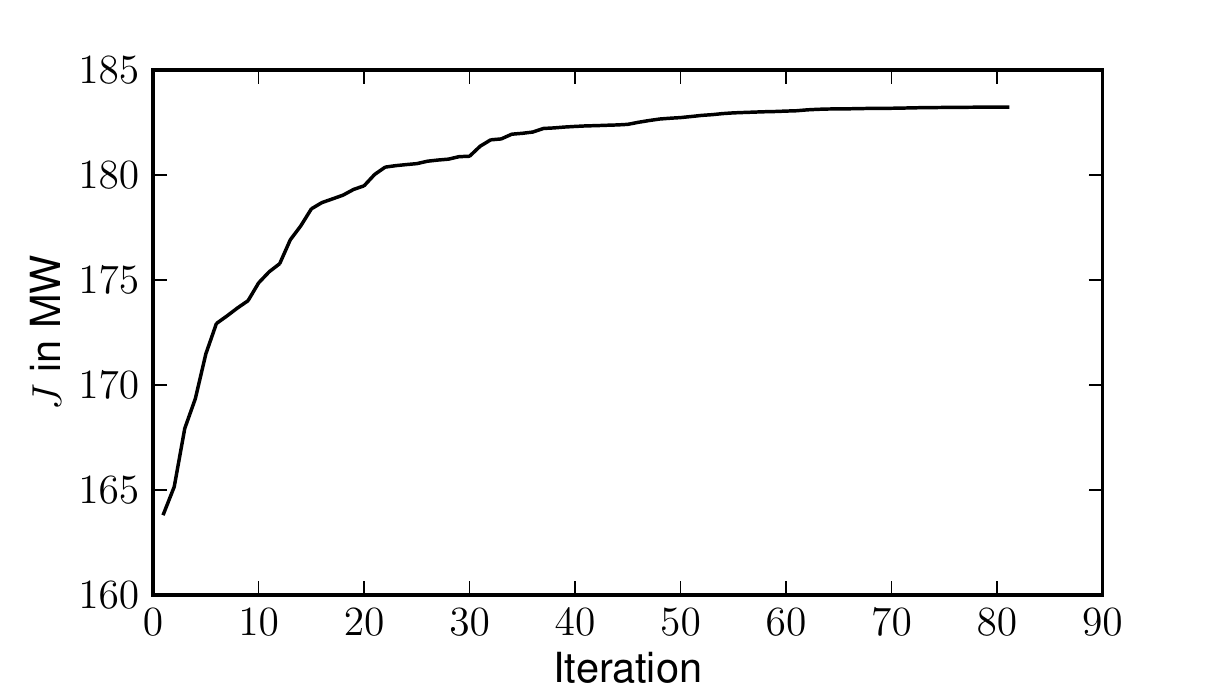}
                \label{fig:orkney_small/iter_plot}
        }
        \\
        \subfloat[Initial turbine positions]{
                \centering
                \includegraphics[width=0.35\textwidth]{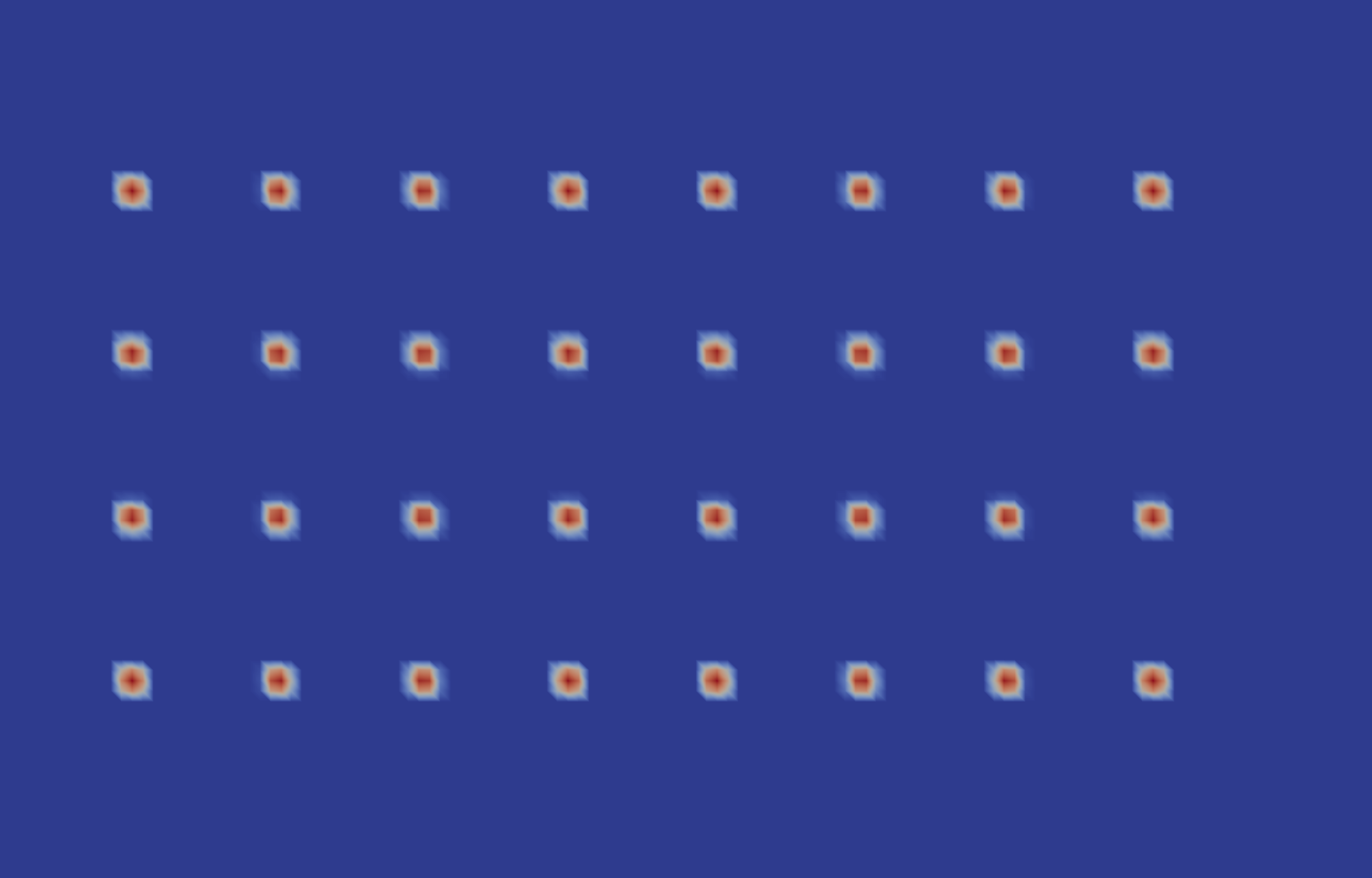}
                \label{fig:initial_layout}
        }
        \hspace{2cm}
        \subfloat[Optimised turbine positions]{
                \centering
                \includegraphics[width=0.35\textwidth]{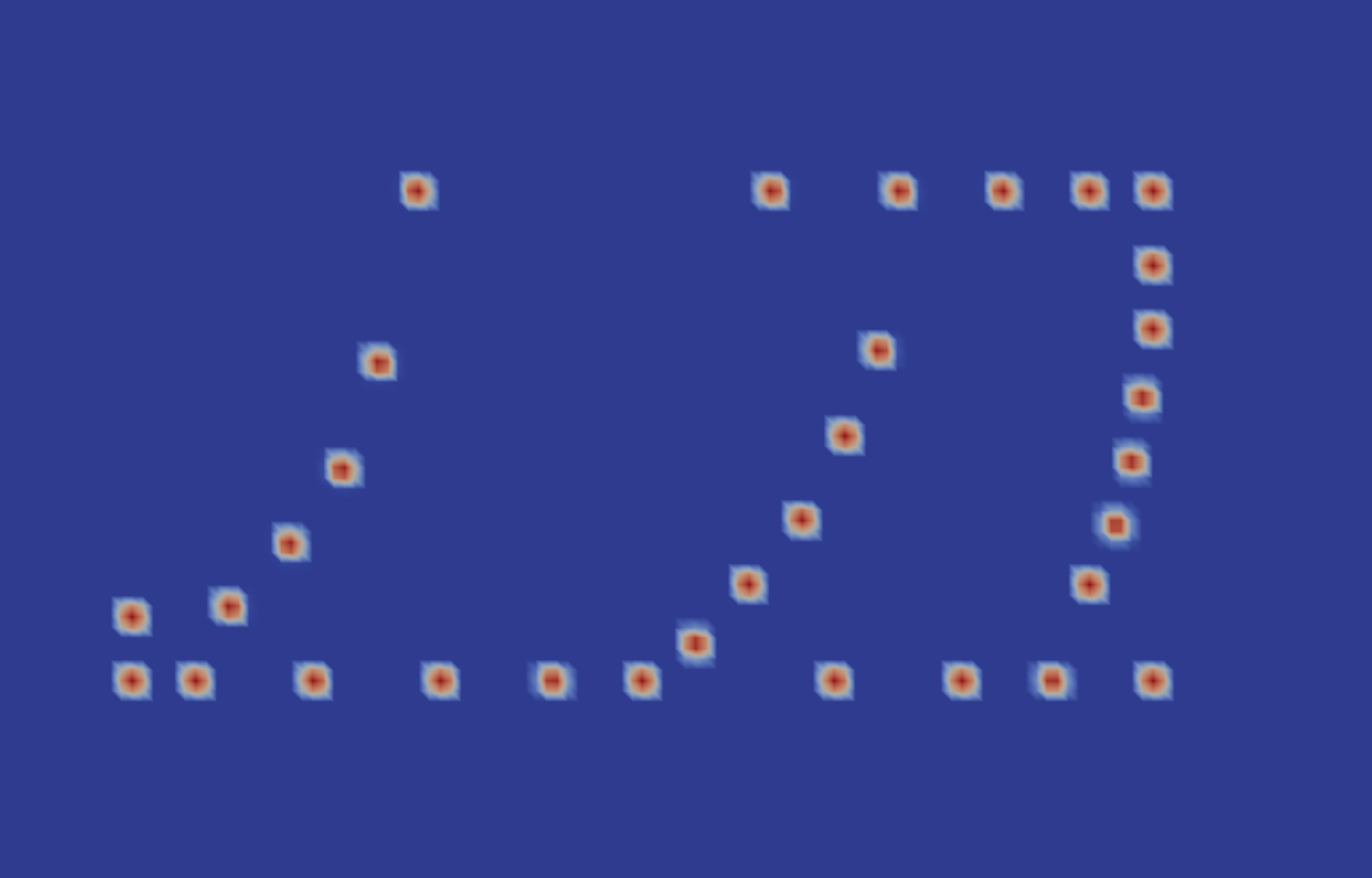}
                \label{fig:turbine_optimisation_final}
        }
        \caption{The solution of the optimal turbine layout problem.}\label{fig:turbine_optimisation}
\end{figure}
\begin{table}
\centering
\begin{tabular}{ccc}
\toprule
       & Runtime (s) & Ratio \\
\midrule
% 4 CPUs
Forward model &  30.41  &     \\
Forward model + adjoint model & 34.22 & 1.125 \\
\bottomrule
\end{tabular}
\caption{Timings for the turbine position gradient calculation. The efficiency of the adjoint is that of the theoretical value of 1.125.}
\label{tab:turbine-timings}
\end{table}

The vector of parameters $m \in \mathbb R^{2N}$ in~\eqref{eq:shallow_water_equations} encodes the $x$ and $y$ positions of $N$ turbines as:
\begin{equation*}
  m = (p_1^x, p_1^y, p_2^x, p_2^y, \dots, p_N^x, p_N^y)^T. 
\end{equation*}
The turbines are parameterised by increased friction located around the turbine centres.
The corresponding friction function $c_t(m)$ is defined as: 
\begin{equation}
 c_t(m)(x, y) \equiv \sum_{i=1}^{N} K \psi_{p_i^x, r}(x) \psi_{p_i^y, r}(y),
\label{eq:turbine_function}
\end{equation}
where $K=21$ is a scaling factor, $r = 40$~m is the extent of the parameterised turbine, and $\psi$ is a smooth bump function with compact support defined as:
\begin{equation*}
\psi_{p, r}(x, y) \equiv
\begin{cases}
  e^{1-1/(1-\|\frac{(x,y)^T - p}{r}\|^2)} &  \mbox{ for } \|\frac{(x,y)^T - p}{r}\| < 1, \\
  0 & \textrm{ otherwise.}
\end{cases}\label{eq:1Dbump_function}
\end{equation*}

The shallow water equations were discretised using the \ptwopone finite-element pair. 
For performance reasons, the function $c_t(m)$ was implemented in Python instead of expressing it as part of the UFL formulation.
Consequently, the dependency on $m$ does not occur explicitly in the UFL form and hence \da cannot automatically compute its derivative.
However, \da is able to automatically compute the derivative of $J$ with respect to $c_t$, and so this problem can be circumvented by
overloading the \texttt{ReducedFunctional} class and manually implementing the final step of the chain rule:
\begin{equation*}
 \nabla \hat J(m) = \frac{\textrm{d} \hat J}{\textrm{d} c_t} \frac{\textrm{d} c_t}{\textrm{d} m}.
\end{equation*}
The first term is automatically computed using \da.
The second term can easily be derived and implemented by hand by differentiating~\eqref{eq:turbine_function}.
Once this \texttt{ReducedFunctional} class was implemented, the optimisation framework could be used as usual.

The example considered here optimises a deployment site near the Orkney Islands in Scotland, where $32$ turbines are to be installed (figure~\ref{fig:orkney_small/domain}).
A constant input flow with $2$ m/s speed is enforced on the left boundary, while the free-surface displacement on the right boundary is set to zero.
A no-normal flow condition is applied on all remaining boundaries. 
The remaining parameters are $H = 50$ m, $\nu = 90\textrm{ m}^2/\textrm{s}$, $g = 9.81\textrm{ m}/\textrm{s}^2$.

The turbines are initially distributed in a structured manner as shown in figure~\ref{fig:initial_layout}. 
The optimisation was performed using the SQP implementation of~\citeN{kraft1994} until the relative reduction of the functional of interest dropped below $10^{-6}$.
Bound constraints ensured that the turbines remained in the site area, which models the fact
that site developers acquire a license for a particular site, and cannot deploy outside it.
Furthermore, a set of inequality constraints were used to enforce a minimum distance of $60$ m between each turbine.

The results are presented in figure~\ref{fig:turbine_optimisation}.
The optimisation algorithm terminated after $81$ iterations. The optimisation increased the total power output of the turbine farm by $12\%$, from an initial value of $164$ MW to $183$ MW.

Table \ref{tab:turbine-timings} compares the runtime of the forward model and the runtime of the gradient calculation.
Both were performed in parallel with 4 CPUs. 
The ratio of forward and adjoint runtimes is close to the theoretical ideal, as expected.  

\subsection{Data assimilation with wetting and drying}
\subsubsection{Introduction}
Wetting and drying processes such as tsunami inundation or the flooding and receding of tides play an important role 
in the study of tsunamis~\cite{kowalik2004}, tidal flats and river estuaries~\cite{zhang2009,xue2010}, and flooding events~\cite{westerink2008,song2010}.
Many algorithms have been proposed for the numerical simulation of wetting and drying processes, 
both for the shallow water equations~(\citeN{medeiros2012} and the references therein) and for the Navier-Stokes equations~\cite{funke2011}.

In this example, we consider a data assimilation problem where the goal is to reconstruct 
the profile of an incoming tsunami from observations of the wet/dry inundation interface.
The tsunami is modelled by the time-dependent nonlinear shallow water equations with the wetting and drying scheme proposed by~\citeN{karna2011}.
The resulting optimisation problem is: 
\begin{equation}
\begin{aligned}
 \min_{m,u,\eta}~&J(\eta) \\
\mbox{subject to }&   \frac{\partial  u}{\partial t} + ( u \cdot \nabla)  u + g \nabla \eta = \frac{c_t(\tilde H)}{\tilde H} \| u\| u && \mbox{on } \Omega \times (0, T], \\
                  & \frac{\partial \tilde H}{\partial t} + \nabla \cdot  (\tilde H  u) = 0 && \mbox{on } \Omega \times (0, T], \\
% u \cdot  n & = 0 && \mbox{on } \partial \Omega_{S} \times [0,T],\\
                  & \tilde H = m && \mbox{on }  \partial \Omega_{D} \times (0, T]. 
% u & =  u_0, \quad \eta = \eta_0 && \mbox{at } \Omega \times \{0\}. 
\end{aligned}\label{eq:modified_shallow_water_with_wetting_and_drying_optimisation_problem}% 
\end{equation} 
where $\Omega \subset \mathbb R^2$ is the spatial domain, $T$ is the final time and $u : \Omega \times (0, T] \to \mathbb R^2$ and $\eta  : \Omega \times (0, T] \to \mathbb R$ are the unknown depth-averaged velocity and free-surface displacement, respectively.
In the classical shallow water equations the total water depth is defined as $H \equiv \eta + h$, where $h :  \Omega \to \mathbb R$ is the static bathymetry; 
in order to account for wetting and drying, \citeN{karna2011} uses a modified total depth definition $\tilde H \equiv f(H)$ instead, where $f$ is a smooth approximation of the maximum operator:
\begin{equation*}
f(H) \equiv \frac{1}{2} \left( \sqrt{H^2 + \alpha^2} + H \right) \approx \max(0, H), \label{eq:wd_function_f}
\end{equation*}
and $\alpha > 0$ controls the accuracy of the approximation.
The remaining parameters in~\eqref{eq:modified_shallow_water_with_wetting_and_drying_optimisation_problem} are the gravitational force $g = 9.81 \textrm{ m}/\textrm{s}^2$ and the friction coefficient in the Ch\'{e}zy-Manning formulation: 
\begin{equation*}
c_t(\tilde H) = \frac{g \mu ^ 2}{\tilde H^{1/3}},
\end{equation*}
where $\mu \in \mathbb R$ is the user specified Manning coefficient.

The boundary conditions are as follows: on the inflow boundary $\partial \Omega_D$ a Dirichlet boundary condition with value $m : (0, T] \to \mathbb R$ is applied, which also acts as the control parameter. 
For simplicity, it is assumed that $m$ varies only in time, i.e. is constant along the boundary. 
On the remaining boundaries, a no-normal flow boundary condition is applied.

The functional of interest $J$ measures the misfit between the observed and the simulated wet/dry interface.
For its formulation, an indicator function is constructed that is $1$ in dry areas and $0$ in wet areas. 
By noting that $\eta \ge h$ in wet areas and $\eta < h$ in dry areas, this indicator function is defined as $\mathcal{H}(\eta - h)$  where 
$\mathcal{H}$ denotes the following smooth approximation of the Heaviside step function:
\begin{equation*}
  \mathcal{H}(x) \equiv \frac{1}{2}\left( \frac{x}{\sqrt{x^2 + \alpha^2}} + 1 \right)\approx
\begin{cases} 
0 & \mbox{if } x < 0, \\ 
1 & \mbox{else},
\end{cases} \label{eq:wd_opt_smooth_heaviside_approx} 
\end{equation*}
where again $\alpha$ controls the smoothness of the approximation.
With that, the functional of interest is defined as:
\begin{equation*}
J(\eta) \equiv \frac{1}{2} \int_0^T \left|\left| \mathcal{H}(\eta - h) - d \right|\right|^2_{L^2(\Omega)} \dt, \label{eq:opt_wd_functional}
\end{equation*}
where $d : \Omega \times (0, T] \to \mathbb R$ denotes the indicator function of the observed wet/dry interface.
While such inverse problems are in general ill-posed and require regularisation, satisfactory numerical results were obtained in this example with no
regularisation term. The implementation of such a regularisation term in the functional would be a trivial modification. 

\subsubsection{Implementation}
The modified shallow water equations in the optimisation problem~\eqref{eq:modified_shallow_water_with_wetting_and_drying_optimisation_problem} are discretised 
with the LBB-stable \ponedgptwo finite element pair in space \cite{cotter2008b}.
A simple upwinding scheme is implemented, which is
obtained by integrating the advection term by parts, replacing the advected
velocity at the inflow facets with the upwind velocity and then integrating
by parts again.
Following \citeN{karna2011}, the resulting equations are then discretised with a second-order Diagonally Implicit Runge-Kutta (DIRK22) scheme in time \cite[\S 2.6]{ascher1997}.

The implementation of problem~\eqref{eq:modified_shallow_water_with_wetting_and_drying_optimisation_problem} 
in the presented optimisation framework was straightforward: 
the control parameters appear directly in the UFL representation of the governing equations,
and hence the framework was applicable without any modifications.

\subsubsection{Reconstruction of the Hokkaido-Nansei-Oki tsunami profile}
This example is motivated by the question of whether it is possible to reconstruct a tsunami profile from satellite observations that record the inundation interface on the coast over time. 

\begin{figure}[bt]
\centering
\subfloat[Bathymetry]{
      \centering
      \includegraphics[width=0.4\textwidth]{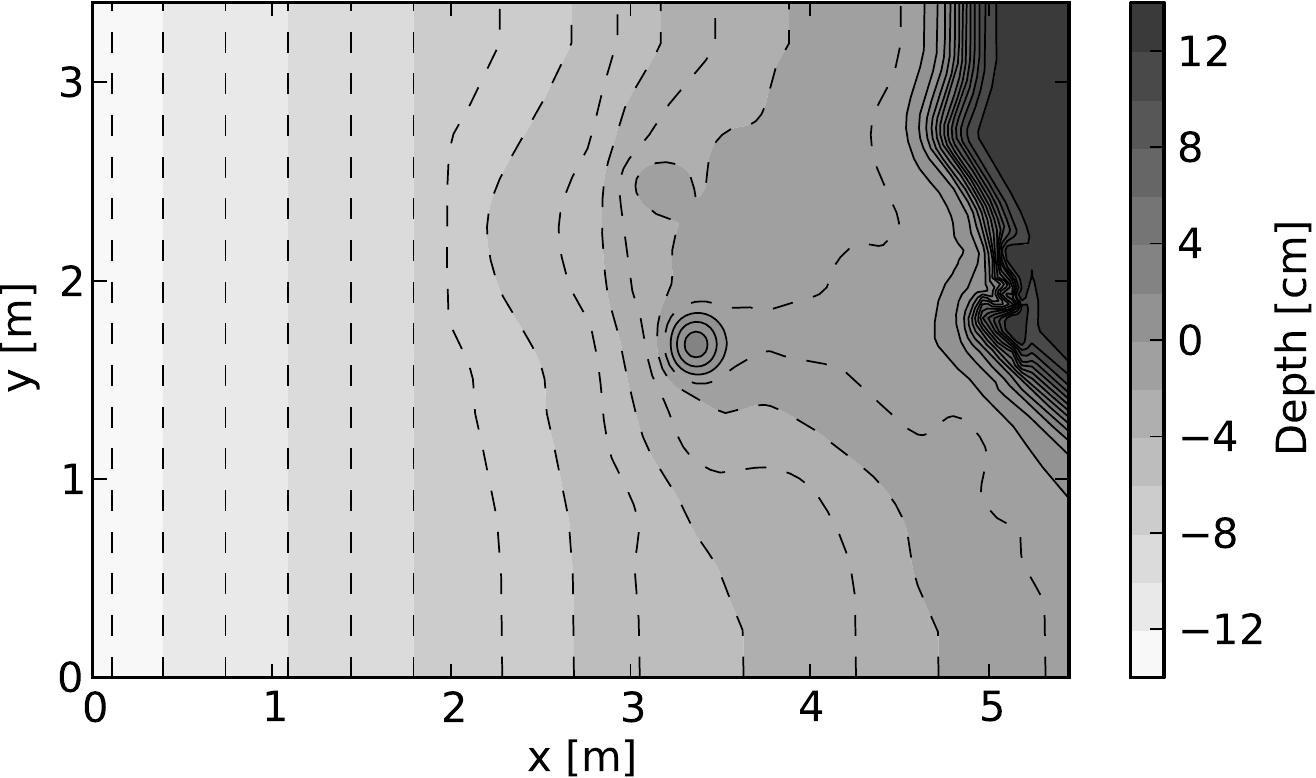}
      \label{fig:wd_optimal_rate_optimal_monai_valley_bathymetry}
     }
  \subfloat[Mesh]{
      \centering
      \includegraphics[width=0.35\textwidth]{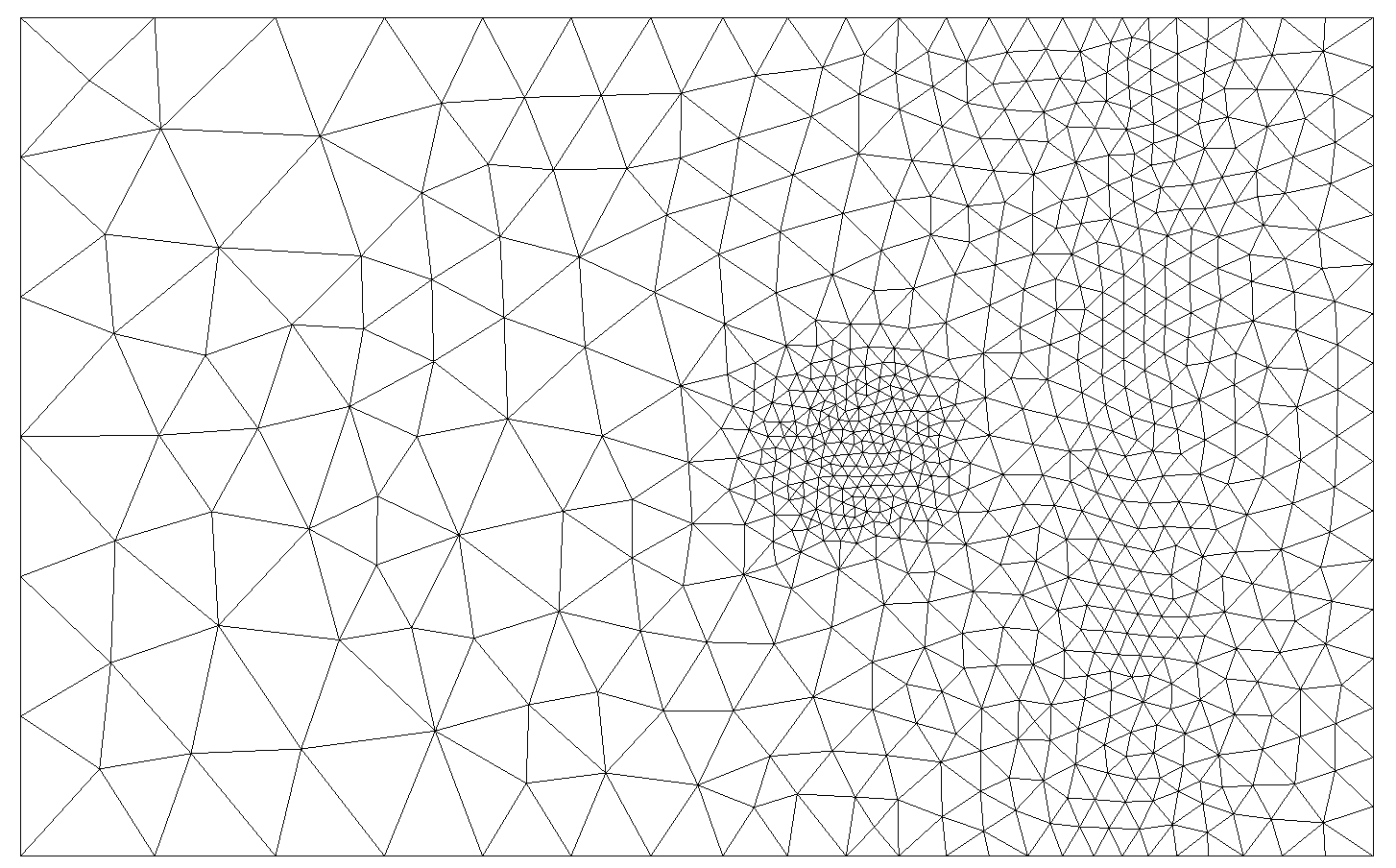}
      \label{fig:wd_optimal_rate_optimal_monai_valley_mesh}
     }
     \caption{The laboratory setup of the Hokkaido-Nansei-Oki tsunami example, based on the $1$:$400$ laboratory experiment of \protect\citeN{matsuyama2001}. The island at the center and the coast on the right are hit by a tsunami coming from the left boundary.} 
\end{figure}

\begin{figure}[t]
\centering
% the original result files can be found in /home/sf1409/Documents/PhD_Simon/thesis/wetting_and_drying_optimisation/wetting_and_drying/hokkaido-nansei-oki_tsunami_final7
%        \subfloat[The wet/dry interface observations after $0$ s, $20$ s, $29$ s, $32$ s, in reading order. The observations are constructed by running the forward problem with the synthetic Dirichlet boundary values.
%         The observations are approximated indicator functions of the wet/dry interface (marked as white lines)]{
%                \centering
%		\includegraphics[width=0.4\textwidth]{hokkaido-nansei-oki_tsunami_final7/results_dg/observation_0.jpg}
%		\includegraphics[width=0.4\textwidth]{hokkaido-nansei-oki_tsunami_final7/results_dg/observation_39.jpg}
%		\includegraphics[width=0.4\textwidth]{hokkaido-nansei-oki_tsunami_final7/results_dg/observation_58.jpg}
%		\includegraphics[width=0.4\textwidth]{hokkaido-nansei-oki_tsunami_final7/results_dg/observation_64.jpg}
%        \label{fig:wetting_drying_tsunami_observations}
%       }
%        \\
\subfloat[The input tsunami profile $m^{\mathrm{opt}}$]{
                \centering
		\includegraphics[width=0.4\textwidth]{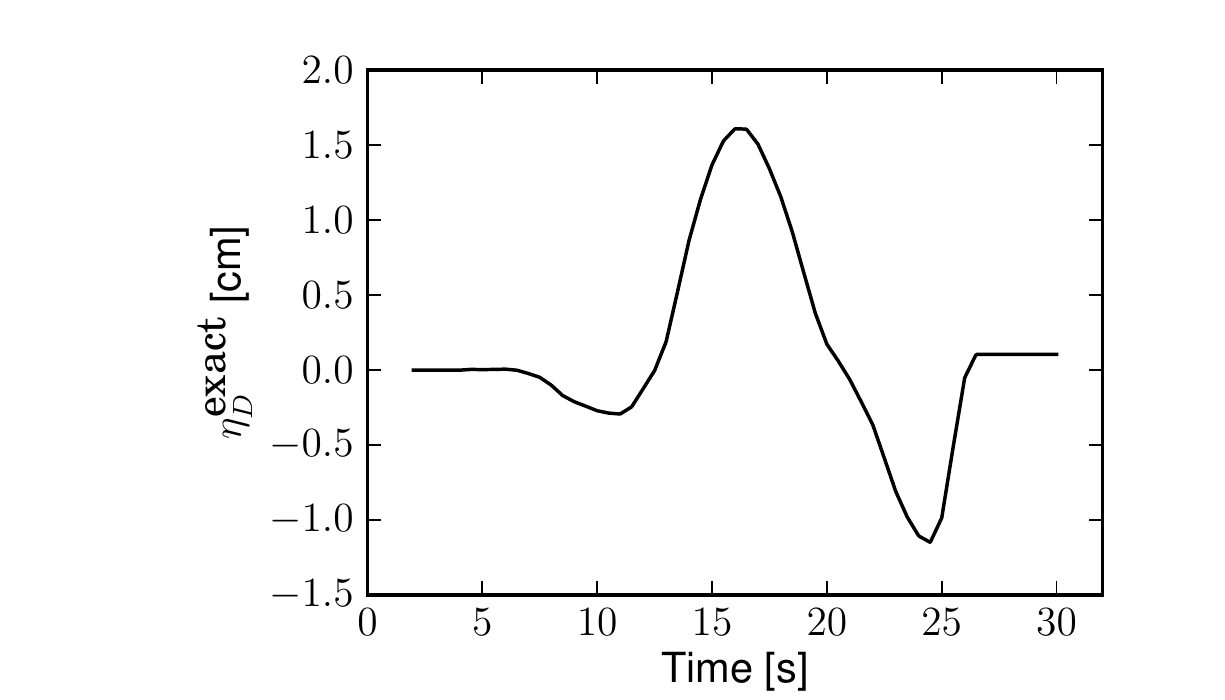}
        \label{fig:hokkaido-nansei-oki_tsunami_controls_optimal}
       }
      \subfloat[The reconstructed tsunami profile $m$]{
                \centering
		\includegraphics[width=0.4\textwidth]{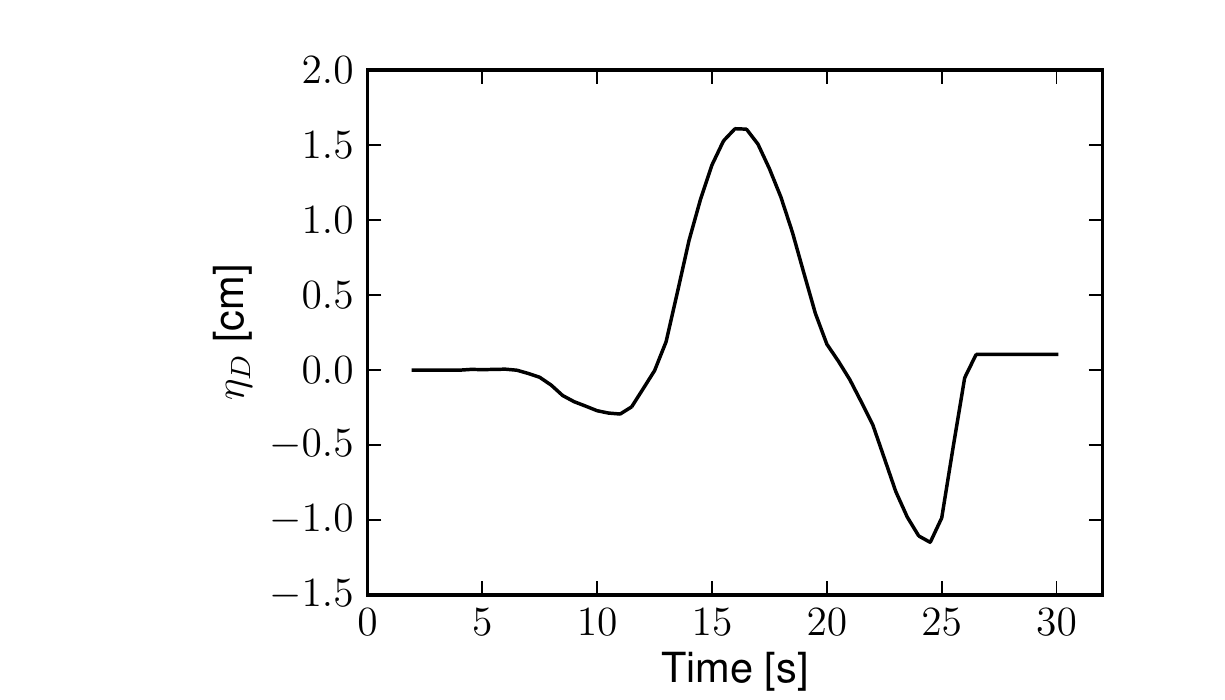}
		\label{fig:wetting_drying_hokkaido-nansei_tsunami_controls_46}
       }\\
       \subfloat[The difference between input and reconstructed tsunami profiles $\left|m - m^{\mathrm{opt}}\right|$]{
                \centering
		% Converged after 103 iterations, reason: the relative decrease of the functional of interest in the last optimisation iteration decreased by less than the machine precision 
		% final Projg value: 1.639D-07
		% 113 functional evaluations
		% maximum absolut error in the Dirichlet control values: 3.09092579656e-09 m
		\includegraphics[width=0.4\textwidth]{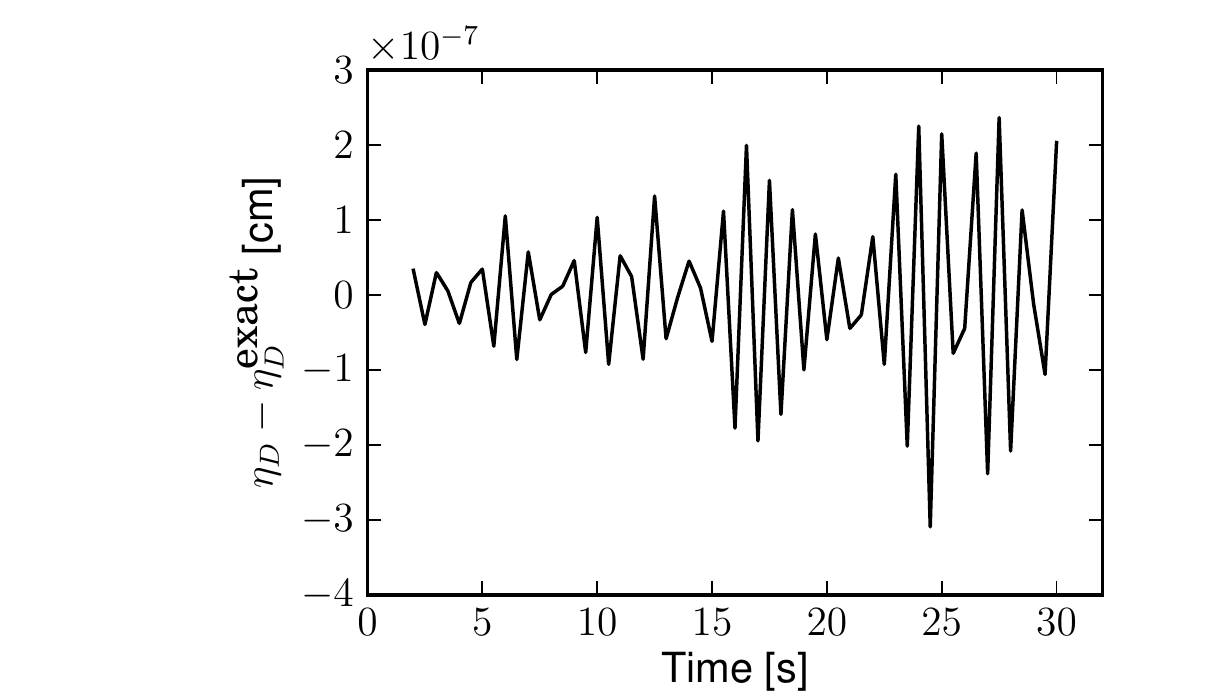}
		\label{fig:hokkaido-nansei-oki_tsunami_controls_errors_46}
       }
       \subfloat[The objective functional $J$]{
                \centering
		\includegraphics[width=0.4\textwidth]{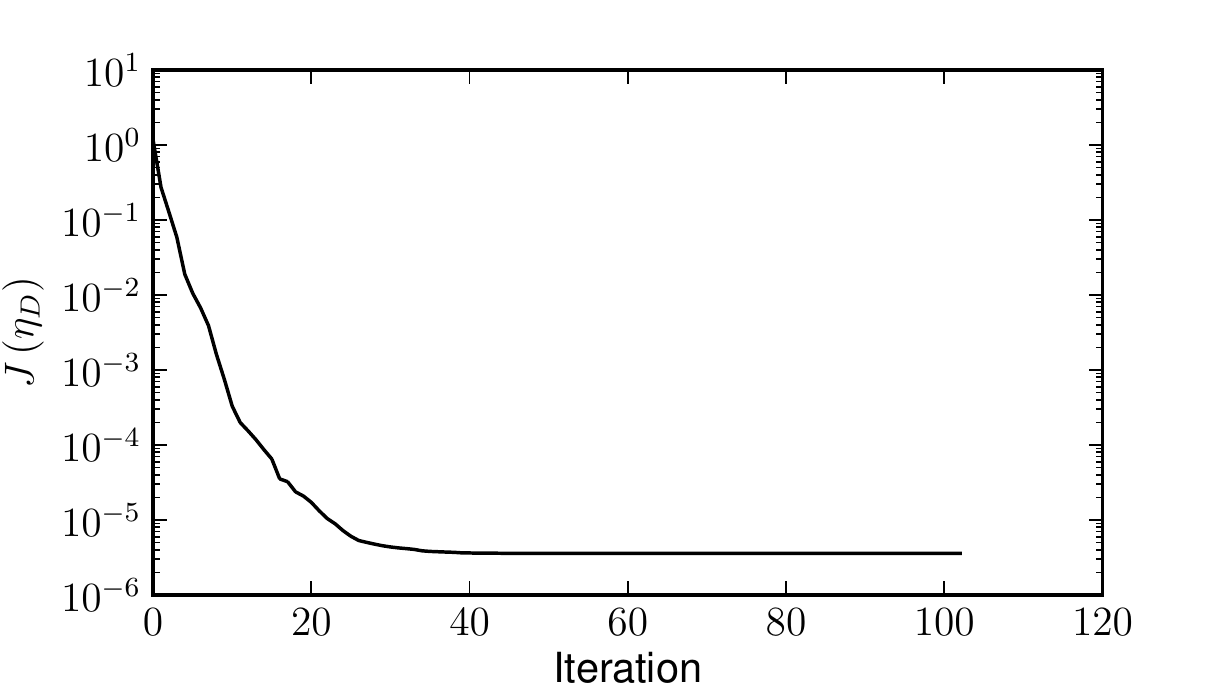}
        \label{fig:wetting_drying_tsunami_iter_plot}
       }
        \caption{Results of the reconstruction of the Hokkaido-Nansei-Oki tsunami profile.
        The initial and final $2$ s of the boundary values are excluded from the reconstruction.}
\label{fig:hokkaido-nansei-oki_tsunami-results}
\end{figure}

The considered event is the Hokkaido-Nansei-Oki tsunami that occurred in 1993 and produced run-up heights of up to $30$~m on Okushiri island, Japan. 
The Central Research Institute for Electric Power Industry (CRIEPI) in Abiko, Japan constructed a $1$:$400$ laboratory scale model of the area around the island \cite{matsuyama2001}. 
Following the setup used in~\citeN{yalciner2011}, we consider a rectangular domain of size $5.448\textrm{ m} \times 3.402\textrm{ m}$, 
with the bathymetry and coastal topography shown in figure \ref{fig:wd_optimal_rate_optimal_monai_valley_bathymetry}. 
It contains an island in the center and coastal regions on the top right of the domain.
The left boundary is the inflow boundary $\partial \Omega_D$, on which a surface elevation profile is enforced that resembles a tsunami (figure \ref{fig:hokkaido-nansei-oki_tsunami_controls_optimal}). The task is to reconstruct this wave profile. 

The domain is discretised with an unstructured mesh consisting of $1,411$ triangular elements with increasing resolution near the inundation areas (figure~\ref{fig:wd_optimal_rate_optimal_monai_valley_mesh}). 
The temporal discretisation uses a timestep of $0.5$~s with a total simulation time of $32$~s. The Manning coefficient was set to $\mu=0.05\textrm{ s}/\textrm{m}^{\frac{1}{3}}$
and a smoothness value of $\alpha = 0.1$ was used.

The observations $d$ are synthetically generated by running the forward model with the wave profile that was used in the laboratory experiment while recording the wet/dry interface. 
This approach of generating the synthetic observation data with the same model that is used for the assimilation is referred to as an ``inverse crime``~\cite{kaipo2005}, 
which renders the optimisation problem less ill-posed than it actually is. 
However, the main purpose of this example is to demonstrate the capabilities of the optimisation framework, and hence this approach is adopted for simplicity.

The optimisation algorithm begins with an initial guess for the Dirichlet boundary values of $0.105$ cm for all timelevels, which corresponds to the final free-surface displacement of the input wave.
The tsunami signal at the boundary is applied $2$~s after the simulation start time.
The boundary condition during the final $2$~s has no impact on the functional,
as the wave does not affect the wet/dry interface before the simulation ends. 
Therefore, the boundary values at the start and end were reset to the correct Dirichlet boundary values and excluded from the optimisation.
Furthermore, a box constraint was used to restrict the minimum and maximum free-surface displacement between $-1.5$ cm and $+2$ cm; 
without these constraints the first optimisation iterations generate unrealistically large Dirichlet boundary values for which the forward model does not converge.

Figure~\ref{fig:hokkaido-nansei-oki_tsunami-results} shows the results of the problem solved with the limited memory BFGS (L-BFGS-B) implementation in SciPy~\cite{scipy}.
After $103$ optimisation iterations ($113$ functional evaluations) the relative decrease of the functional of interest fell below machine precision and the algorithm terminated. 
The incoming wave was reconstructed to within an absolute error of $3.91 \times 10^{-7}$ cm (figure~\ref{fig:hokkaido-nansei-oki_tsunami_controls_errors_46}),
which corresponds to a relative error of less than $3\times10^{-5}$\% of the incoming wave height.

\begin{table}
\centering
\begin{tabular}{ccc}
\toprule
       & Runtime (s) & Ratio \\
\midrule
%Smallest runtime for forward model:  2025.95630121
%Smallest runtime for forward + adjoint model:  3145.88366008
%Ratio:  1.55278949412
Forward model & 2026 &     \\
Forward model + adjoint model & 3146 & 1.55 \\
\bottomrule
\end{tabular}
\caption{Timings for the tsunami reconstruction gradient calculation. 
The efficiency of the adjoint approaches the theoretical ideal value of 1.5.}
\label{tab:tsunami-timings}
\end{table}

Table \ref{tab:tsunami-timings} lists the runtimes of the forward model and the gradient calculation.
The nonlinear equations of the forward model are typically solved with 2 Newton iterations, 
which suggests an optimal runtime ratio of 1.5. 
The measured value is close to this theoretical value and confirms the relative efficiency of the adjoint model implementation. 

%%%%%%%%%%%%%%%%%%%%%%%%%%%%%%%%%%%%%%%%%%%%%%%%%%%%%%%%%%%%%%%%%%%%%%%%%%%%%%%%%%%%%

\section{Summary}\label{sec:summary}
In this paper we present a new framework for rapidly defining and solving PDE-constrained optimisation problems.
The framework exploits the \fenics system, \da, and established optimisation algorithms to allow the user
to specify the discretised optimisation problem in a high-level language that resembles the mathematical notation.

The core idea of the implementation is to perform all required tasks of the optimisation process using the tape of the forward model that is recorded by \da 
(analogous to the AD concept of a tape).
This includes the evaluation of the objective functional by replaying the forward model,
computing the functional gradient by deriving and solving the adjoint problem, 
and modifying the tape in order to incorporate parameter updates.
While this paper applies the idea to the \da system, the same approach is applicable to the operator-overloading class of AD tools that build a tape at runtime.

As demonstrated, this approach reduces the required user input to a minimum: once the forward model has been implemented, 
only a handful of lines of code are required to specify the optimisation problem.
It applies naturally to both linear and nonlinear as well as to both steady and time-dependent governing PDEs.
Furthermore, the user has the choice of a variety of gradient-free and gradient-based methods.
General equality and inequality control constraints can be applied.

In this paper, the reduced formulation is used for solving the optimisation problem. 
For cases where quasi-Newton methods applied to the reduced formulation are insufficient, 
the framework provides all ingredients necessary for more sophisticated approaches.
Therefore, this framework is also of interest for the development of such advanced optimisation algorithms: by implementing an algorithm
in the framework once, it is immediately applicable to optimisation problems across science and engineering.
Future work includes the automation of shape optimisation, the development of the oneshot approach, multigrid optimisation techniques, and the exploitation of reduced-order
modelling in optimisation.

\bibliographystyle{acmsmall}
\bibliography{literature}
\end{document}